\documentclass[12pt,preprint]{aastex}
\shorttitle{}
\shortauthors{}
\begin{document}
\baselineskip 19.pt
\title{Dynamics of Dust Particles Released from Oort Cloud Comets
and Their Contribution to Radar Meteors}
\author{David Nesvorn\'y$^1$, David Vokrouhlick\'y$^{1,2}$, Petr Pokorn\'y$^{1,2}$, 
Diego Janches$^3$} 
\affil{(1) Department of Space Studies, Southwest Research Institute,  
1050 Walnut St., Suite~300, Boulder, Colorado 80302, USA}
\affil{(2) Institute of Astronomy, Charles University, \\
V Hole\v{s}ovi\v{c}k\'ach 2, CZ-18000, Prague 8, Czech Republic}
\affil{(3) Space Weather Laboratory, Code 674, GSFC/NASA, \\Greenbelt, 
MD 20771, USA}

\begin{abstract}
The Oort Cloud Comets (OCCs), exemplified by the Great Comet of 1997 (Hale-Bopp), are 
occasional visitors from the heatless periphery of the solar system. Previous works hypothesized that 
a great majority of OCCs must physically disrupt after one or two passages through the inner solar 
system, where strong thermal gradients can cause phase transitions or volatile pressure buildup.
Here we study the fate of small debris particles produced by OCC disruptions to determine whether 
the imprints of a hypothetical population of OCC meteoroids can be found in the existing meteor 
radar data. We find that OCC particles with diameters $D \lesssim 10$ $\mu$m are blown out from the 
solar system by radiation pressure, while those with $D \gtrsim 1$ mm have a very low Earth-impact 
probability. The intermediate particle sizes, $D\sim100$ $\mu$m, represent a 
sweet spot. About 1\% of these particles orbitally evolve by Poynting-Robertson drag to reach orbits 
with semimajor axis $a \sim 1$ AU. They are expected to produce meteors with radiants near the apex 
of the Earth's orbital motion. We find that the model distributions of their impact speeds and orbits 
provide a good match to radar observations of apex meteors, except for the eccentricity 
distribution, which is more skewed toward $e\sim 1$ in our model. Finally, we propose an explanation 
for the long-standing problem in meteor science related to the relative strength of apex and 
helion/antihelion sources. As we show in detail, the observed trend, with the apex meteors being more 
prominent in observations of highly sensitive radars, can be related to orbital dynamics of 
particles released on the long-period orbits. 
\end{abstract}
\section{Introduction}
The Oort cloud is a roughly spherical cloud of comets (Oort 1950), which surrounds the solar system
and extends to heliocentric distances larger than 100,000 AU. The Oort cloud is currently feeding 
comets into the inner solar system at a rate of about 12 comets with $q<3$~AU per year with an 
active comet absolute magnitude $H_{10}<10.9$ (Wiegert \& Tremaine 1999; $q$ is the 
perihelion distance and $H_{10}$ is a distance-independent measure of the active comet brightness 
that includes the coma). The Oort Cloud Comets (hereafter OCCs) can be divided into two populations 
based on their dynamical histories: (1) dynamically new OCCs, which are on their first passage through
the inner solar system and typically have $a\gtrsim10,000$ AU; and (2) returning OCCs which have 
previously passed through the inner solar system and typically have $a<10,000$ AU.\footnote{See 
Dybczynski (2001) for an alternative definition.}

The dynamical models of the orbital evolution of new OCCs into returning OCCs predict many times
more returning comets than are observed (Wiegert \& Tremaine 1999). This is the so-called ``fading
problem'', which is thought to be related to the physical evolution of OCCs. To resolve this problem, 
Levison et al. (2002, hereafter L02) proposed that OCCs must physically disrupt as they evolve inward 
from the Oort cloud. Specifically, L02 estimated that, when a OCC becomes inactive, it has only 
$\sim$1\% chance of becoming dormant, and $\sim$99\% chance of being disrupted. If more OCCs would 
become dormant, L02 argued, the modern surveys of near-Earth objects would discover a  
greater number of dormant OCCs passing through perihelion each year than they do. Strong thermal 
gradients, phase transitions, and volatile pressure buildup experienced by OCCs during their 
approaches to the Sun are thought to be responsible for disruptions. 

If these results are correct, the disrupted OCCs must be a prodigious source of dust particles 
and larger fragments that may further disintegrate. In particular, the dust production rate from 
OCC disruptions should be vastly larger than that of active OCCs. On the other hand, the smallest 
dust particles produced in these disruption events may be lost from the Solar System due to the 
effects of radiation pressure, while the large fragments should be dispersed over enormous radial 
distances. It is therefore not clear whether the disrupted OCCs can supply a significant amount 
of material to the inner zodiacal cloud, and whether they could represent a significant source of 
interplanetary dust particles accreted by the Earth. 

Given their large speeds relative to the Earth, the OCC particles plunging into the upper atmosphere
could produce meteor phenomena and be detected by optical and radar meteor surveys (e.g., see Steel 
(1996) for a review). Here we consider the radar meteors. The modern meteor radar surveys 
produced vast datasets including millions of high quality orbits covering both the northern 
and southern hemispheres (e.g., Jones \& Brown 1993, Taylor \& Elford 1998, Galligan \& Baggaley 
2004, 2005, Janches et al. 2003, Janches \& Chau 2005, Chau et al. 2007, Campbell-Brown 2008). 
Moreover, the sensitivity of meteor observations broke new grounds with routine measurements of
meteor echoes using HPLA\footnote{High Power and Large Aperture.} radars such as the Arecibo (AO) 
radar, which is capable of detecting $\sim$50-$\mu$m particles down to $\sim$20 km s$^{-1}$ (e.g., 
Janches et al. 2003, 2006, Fentzke et al. 2009). It is natural to ask whether some of these 
observations can be linked to the particle populations from disrupted OCCs.

Meteors are produced by small interplanetary particles, also known as the {\it meteoroids}, that
interact with air molecules upon atmospheric entry. Based on meteor 
data, the meteoroids can be divided into two groups: sporadic meteoroids and meteoroid streams. The 
meteoroid streams are prominent concentrations of particles with similar orbits (Whipple \& Gossner 1949, 
Whipple 1951). 
They are thought to be produced by particles released by active and recently ($<$ few thousand years ago) disrupted 
comets (e.g., Jenniskens 2008). Sporadic meteoroids are those particles that have evolved significantly 
from their parent body so that they are no longer easily linked to that parent, or to other meteoroids 
from the same parent. Notably, the time-integrated flux of meteors at Earth is dominated by about a factor 
of $\sim$10 by sporadics (Jones \& Brown 1993).

The radiant distribution of sporadic meteors shows several concentrations on the sky known as the 
helion/antihelion, north/south apex, and north/south toroidal sources (e.g., Younger et al. 2009, and 
the references therein). Wiegert et al. (2009; hereafter W09) developed a dynamical model to explain these 
concentrations. Their main results concern the prominent helion/antihelion sources for which the 
particles released by Jupiter-Family Comets (JFCs) such as 2P/Encke provide the best match, in 
agreement with previous studies (e.g., Jones et al. 2001). As for the north/south apex source,  W09 pointed 
out the potential importance of retrograde Halley-type comets (HTCs) such as 55P/Tempel-Tuttle (or an 
orbitally similar lost comet). The case for the retrograde HTC particles is compelling, 
because three prominent retrograde HTCs, namely 1P/Halley, 55P/Tempel-Tuttle and 109P/Swift-Tuttle,
all have associated streams, known as Orionids/$\eta$ Aquarids, Leonids and Perseids, respectively. 
The sporadic meteoroids with the north/south apex radiants can thus plausibly be a dynamically old component
of HTC particles.

Here we consider the possibility that at least some part of the meteoroid complex is produced 
by disrupting OCCs (L02, see also Jones et al. 2001). 
We study the effects of radiation pressure on particles released from the 
highly-eccentric OCC orbits, and their dynamical evolution under gravitational perturbations from planets and 
Poynting-Robertson (P-R) drag (see Section 2 for our model). We show that a significant contribution of 
OCC particles to the inner zodiacal cloud and meteor record is somewhat problematic, because most small 
OCC particles are blown out from the solar system by radiation pressure, while most large ones get scattered 
by planets and never make it into the inner solar system (Section 3). Still, we find that there is a 
sweet spot at particle sizes $\sim$100-300 $\mu$m. Our modeling work shows that the orbits and impact 
speeds of these intermediate-size OCC particles can match those derived from the meteor radar data for 
apex meteoroids. Furthermore, we find that the preponderance of fast apex meteors in HPLA radar observations
(e.g., AO, ALTAIR, Jicamarca) can be linked to the competing effects of P-R drag and Jupiter perturbations, 
which act as a size filter on populations of the long-period meteoroids (Section 4).   
\section{Model}
We studied the following sequence of events: 
(1) particles of different sizes were released from OCCs (Section 2.1), (2) their orbits evolved under 
the influence of gravitational and radiation forces (Section 2.2), (3) some particles were thermally
or collisionally destroyed (Section 2.3), and (4) a small fraction of the initial particle population 
was accreted by Earth, producing meteors (Section 2.4). We describe our model for 
(1)-(4) below. 

\subsection{Initial Orbits}
According to Francis (2005), OCCs have ${\rm d}N(q) \propto (1 + \sqrt{q})\, 
{\rm d}q$ for $q<2$ AU. For $q > 2$ AU, Francis' result predicts ${\rm d}N(q)$ being flat or declining
while we would expect the perihelion distribution to increase with $q$. It probably just 
shows that the distribution is not well constrained for $q > 2$ AU. We used ${\rm d}N(q) \propto 2.41 
(q/2)^\gamma\, {\rm d}q$ for $q>2$ AU, with $0\leq\gamma\leq1$. The initial values of $q$ 
in our numerical integrations were set to be uniformly random between 0 and 5 AU, because particles starting 
with $q>5$ AU do not reach 1 AU (see Section 3.1), where they could contribute to the Earth impact record. 
The results for different ${\rm d}N(q)$ were obtained by assigning the appropriate weight to 
particles starting with different $q$'s. 

Upon its release from a larger object a small particle will feel the effects of radiation pressure. 
These effects can be best described by replacing the mass of the Sun, $m_\odot$, by $m_\odot(1-\beta)$, 
with $\beta$ given by
\begin{equation} \beta = 5.7\times10^{-5}\ {Q_{\rm pr} \over \rho s}\ , \label{beta} \end{equation} 
where radius $s$ and density $\rho$ of the particle are in cgs units. Pressure coefficient $Q_{\rm pr}$ 
can be determined using the Mie theory (Burns et al. 1979). We set $Q_{\rm pr} = 1$, which corresponds 
to the geometrical optics limit, where $s$ is much larger than the incident-light wavelength. We used 
particles with $D=2s=10$, 30, 100, 300, 1000 $\mu$m, which should cover the interesting range of sizes,
and $\rho=2.0$ g cm$^{-3}$. 

For large eccentricity $e$ of the parent object and/or for large $\beta$, the released particle may 
become unbound and escape to interstellar space. To stay bound, the heliocentric distance, $R$, 
of the released particle must fulfill the following condition (e.g., Kresak 1976, Liou et al. 1999): 
\begin{equation}
R>R^*=2 \beta a\ .
\label{rstar}
\end{equation}
This condition shows that all particles with $\beta$ released at the orbit's perihelion will be 
removed, if $2\beta > 1-e$. The new OCCs have $1-e \lesssim 10^{-4}$. It follows that particles 
produced by a new OCC near its perihelion will become unbound for sizes up to $D\sim1$ cm. 
The usual near-perihelion activity of OCCs therefore cannot be a major source of small dust 
particles in the inner solar system.

Motivated by the L02 results, we now consider OCCs disruptions. Interestingly, observations of 
the disruption events of comets show that there does not seem to be any correlation between the time 
of disruption and the orbital phase of the parent object. Many comets were seen to disrupt 
(or suffer outburst/splitting events) at large heliocentric distances. For example, 174P/Echeclus
showed an outburst with $R\approx13$ AU, more than 6 AU beyond its perihelion distance (Choi et al. 
2006). It may therefore be possible that OCCs could disrupt at relatively large $R$ and produce 
particles that, according to Eq. (\ref{rstar}), will stay on bound orbits.

We release particles with $R>R^*$ in our model. For example, a $D=100$-$\mu$m particle with 
$\rho=2.0$~g~cm$^{-3}$ ejected from the parent comet with $a=10^3$ AU will have 
$\beta \simeq 0.006$ and $R^*=12$~AU. We thus release these particles with $R>12$ AU. In addition, 
we only study particles ejected from orbits similar to those of the {\it returning} OCCs with 
$a\sim10^3$ AU. We do not consider orbits with $a\gtrsim10^4$ AU, corresponding to the Oort 
spike\footnote{The semimajor axis values of most OCCs are $10^4 \lesssim a \lesssim 5\times10^4$ AU, which 
is known as the Oort spike (e.g., Wiegert \& Tremaine 1999). Comets in the spike are mostly 
dynamically new comets, on their first passage into the inner planetary system from the Oort cloud.
A comet that passes through the planetary system receives a gravitational kick from the planets.
The typical energy kick, $\Delta x$, depends strongly on the perihelion distance of the comet's orbit. 
According to Wiegert \& Tremaine (1999), $\Delta x \sim 10^{-3}$ AU$^{-1}$ for $q\lesssim6$ AU, while 
comets in the Oort spike have $x=1/a\lesssim10^{-4}$ AU$^{-1}$. Depending on the sign of the kick,
they will either leave the planetary system on unbound orbit, never to return, or be thrown onto 
a more tightly bound orbit with $a\lesssim10^3$ AU.}, 
because we believe it unlikely that disruptions could happen at the very large heliocentric distance 
implied by Eq. (\ref{rstar}) for $a\gtrsim10^4$ AU. For example, a $D=300$-$\mu$m particle with 
$\rho=2.0$~g~cm$^{-3}$, released from a parent orbit with $a=10^4$ AU, would become unbound, unless 
$R>36$~AU. 

The particle populations studied here have bound initial orbits. They represent only a fraction of 
all particles released from OCCs. This fraction, denoted $f_0$ in Section 4, is difficult to 
estimate, because we do not have a detailed understanding of the processes, and their 
dependence on $R$, that govern comet disruptions. We will return to this issue in Section 4.
The initial distribution of orbital inclination vectors was set to be isotropic. To 
simplify things, we neglected the ejection velocities of dust particles from their parent objects 
(see Jones et al. 2001) and assumed that they will initially follow the parent comet's orbit modified 
by radiation pressure.  

For reference, we also followed meteoroids from 1P/Halley, 2P/Encke and 55P/Tempel-Tuttle. These comets
were suggested to be important sources of the sporadic meteors by W09. The comet's orbits 
were obtained from JPL Horizons site. Particles of different sizes were released from the parent
orbits and tracked into future. We applied the same procedures/criteria to them that we used for the 
OCC particles.    
\subsection{Orbit Integration} 
The orbits of small particles in the interplanetary space are subject to gravitational perturbations 
of planets and radiation forces (Robertson 1937, Burns et al. 1979). The acceleration $\vec{F}$ due to 
radiation forces is 
\begin{equation} \vec{F} = \beta G {m_\odot \over R^2} \left[ \left (1 - {\dot{R} \over 
c} \right) {\vec{R} \over R}  - {\vec{V} \over c} \right] \ , \label{acc} \end{equation}  
where $\vec{R}$ is the heliocentric position vector of particle, $\vec{V}$ is its velocity,  
$G$ is the gravitational constant, $m_\odot$ is the mass of the Sun, $c$ is the speed  of 
light, and $\dot{R} = {\rm d}R/{\rm d}t$. The acceleration (\ref{acc}) consists of the 
radiation pressure and the velocity-dependent P-R term. Parameter $\beta$ is related to the 
radiation pressure coefficient $Q_{\rm pr}$ by Eq. (\ref{beta}). 

The particle orbits were numerically integrated with the {\tt swift\_rmvs3} code (Levison \&
Duncan 1994), which is an efficient implementation of the Wisdom-Holman map (Wisdom \& Holman 1991) 
and which, in addition, can deal with close encounters between particles and planets. The radiation 
pressure and drag forces were inserted into the Keplerian and kick parts of the integrator, 
respectively. The change to the Keplerian part was trivially done by substituting  $m_\odot$ 
by $m_\odot(1-\beta)$. The {\tt swift\_rmvs3} integrator is stable even for near-parabolic 
orbits, and thus well suited for the integrations that we carried out here. 

The code tracks the orbital evolution of a particle that revolves around the Sun and is 
subject to the gravitational perturbations of seven planets (Venus to Neptune; the mass 
of Mercury was added to the Sun) until the particle impacts a planet, is ejected from the Solar 
System or evolves to within 0.05~AU from the Sun. We removed particles that evolved to $R<0.05$~AU, 
because the orbital period for $R<0.05$~AU was not properly resolved by our 1-day integration 
timestep. 

Several thousand 
particles were followed for each $D$. Their orbital elements were defined with respect to the 
barycenter of the solar system. The barycentric elements are similar to the heliocentric
elements for $R<5$ AU, but differ for large $R$, where the Sun's orbital speed about the 
barycenter of the solar system is not negligible relative to the orbital speed of a 
particle.

All orbits were followed from the present epoch into the future. Each particle's orbital elements were 
stored at $10^3$ yr intervals. We used the output to construct a {\it steady state 
distribution} of OCC particles in the inner solar system. This approach differs from that of Nesvorn\'y 
et al. (2006) and W09, who started particles at many different past epochs and 
used these integrations to determine the {\it present distribution} of particles. The two distributions 
are expected to be slightly different expressing mainly the difference between the present configuration 
of planets, with each planet having a specific secular phase, and the time-averaged system, where all 
phases are mixed. Since this difference is small, however, we can use the steady state distribution, 
which is easier to obtain, as a reasonable approximation.
\subsection{Physical Effects}
Solar system micrometeoroids can be destroyed by collisions with other particles and by solar heating 
that can lead to sublimation and vaporisation of minerals. Here we describe how we parametrize these 
processes in our model. 

\subsubsection{Thermal Destruction}
Thermal alteration of grains in the interplanetary grains is a complex process. The OCC particles
evolving into the inner solar system will first loose their volatile ices, which will rapidly sublimate
once the grains are heated to a critical temperature. We do not model the volatile loss in detail.
Instead, we crudely assume that the grains have lost $\sim$50\% of their mass/volume when reaching 
$R\lesssim5$ AU. We do not include the orbital effects of mass loss in orbital modeling, because
it should produce only a relatively small perturbation on orbits for large particles that we consider 
here. The remaining grains will be primarily composed from amorphous silicates and will survive 
down to very small $R$. 

According to Duschl et al. (1996), silicates are thermally altered at temperatures $T\sim900$-1600 K, 
and start to vaporise for $T>1600$ K. As an example of thermal alteration, Kasuga et al. (2006, see 
also \v{C}apek \& Borovi\v{c}ka 2009) studied 
the thermal desorption of alkali minerals and concluded that micrometeoroids should show evidence 
of thermal desorption of metals, Na in particular, for $q<0.1$ AU. Following Moro-Mart\'{\i}n \& Malhotra
(2002), Kessler-Silacci et al. (2007) and others, we adopt a simple criterion for the silicate grain
destruction. We will assume that they are destroyed when the grain temperature reaches $T=1500$ K. 

The temperature of a small, fast spinning grain in interplanetary space is set by an equilibrium 
between the absorbed and re-radiated energy fluxes. While the absorbed flux is a simple function of
the particle's size, albedo and heliocentric distance, the re-radiated flux depends on the particle's 
emissivity, which in turn is a function of particle's size, shape and material properties. Using 
the optical constants of amorphous pyroxene of approximately cosmic composition (Henning \& 
Mutschke 1997), we find that a dark $D\gtrsim100$-$\mu$m grain at $R$ has the equilibrium temperature 
within 10 K of a black body, $T(R)\simeq 275/\sqrt{R}$ K. According to our simple destruction 
criterion, $T(R)>1500$ K, the silicate grains should thus be removed when reaching $R\lesssim0.03$ AU. 
On the other hand, the smallest particles considered in this work, $D=10$ $\mu$m, will reach
$T(R)=1500$ K for $R\simeq 0.05$ AU. Thus, we opted for using a very simple (and conservative) 
criterion where particles of all sizes were destroyed, and not considered for statistics, if they 
ever reached $R \leq 0.05$ AU. Note that, by design, this limit is the same as the one imposed by the 
integration timestep (Section 2.2).          

\subsubsection{Disruptive Collisions}
The collisional lifetime of meteoroids, $\tau_{\rm coll}$, was taken from Gr\"un et~al. 
(1985; hereafter G85). It was assumed to be a function of particle mass, $m$, and orbital parameters,
mainly $a$ and $e$. We neglected the effect of orbital inclination on $\tau_{\rm coll}$, 
because the results discussed in Steel \& Elford (1986) suggest that the inclinations 
should affect $\tau_{\rm coll}$ only up to a factor of $\sim$2-5, which is not overly significant
in the context of our work. We assumed that the mass and orbital dependencies of $\tau_{\rm coll}$ 
can be decoupled, so that
\begin{equation}
 \tau_{\rm coll}\left(m,a,e\right) = \Phi\left(m\right)
  \Psi\left(a,e\right) \; , \label{col1}
\end{equation}
where $\Phi(m)$ and $\Psi(a,e)$ are discussed below.

\begin{figure}
\epsscale{0.8}
\plotone{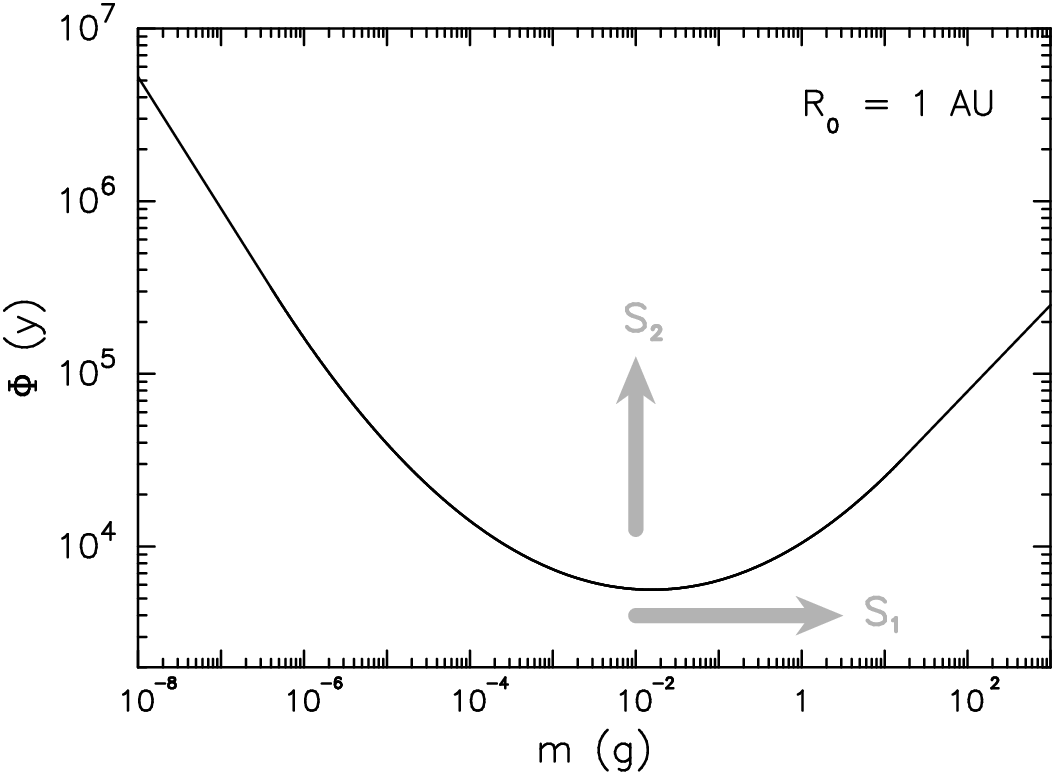}
\caption{The mass dependence, $\Phi(m)$, of the adopted model for the 
collisional lifetime of particles. The plot shows $\Phi(m)$ for a particle on the
circular orbit at the reference heliocentric distance $R_0=1$ AU. Two parameters 
of the collisional model, $S_1$ and $S_2$, were used to test the sensitivity of 
our results to modifications of $\Phi(m)$.}
\label{gf}
\end{figure}

As for $\Phi(m)$, we used the G85's model based on measurements of various spacecraft and 
Earth-based detectors. We found that the model can be approximated by a simple 
empirical fit. Specifically, between $10^{-6.4}$~g and $10^{1.2}$~g, we adopted the quadratic 
function
\begin{equation}
  \log \Phi\left(m\right) = c_2 \left(\log m\right)^2 +
   c_1 \left(\log m\right) + c_0 \; , \label{col2}
\end{equation}
with $(c_0,c_1,c_2)=(4.021,0.300,0.083)$, where the values of $c_0$, $c_1$ and $c_2$ were set 
to fit the G85's collision lifetime for circular orbits at the reference distance $R_0=1$~AU. 
A linear relation between $\log \Phi\left(m\right)$ and $\log m$ was used outside the quoted 
mass range to approximate the G85's model down to $m = 10^{-8}$~g. $\Phi(m)$ has a minimum 
for $m \simeq 0.01$~g, corresponding to $s\simeq 1$ mm for $\rho=2$ g cm$^{-3}$ (Fig.~\ref{gf}). 
The collisional lifetime of $\sim$1 mm particles in the G85's model is very short, roughly 
5,000 yr at 1 AU.

$\Psi(a,e)$ is assumed to drop as a power-law with $R$. From Eq. (18) in G85 we have
\begin{equation}
 \Psi =   \left(\frac{R}{R_0}\right)^\alpha
  \frac{v\left(R\right)}{v_{\rm circ}\left(R\right)} \; , \label{col3}
\end{equation}
where $v(R)$ and $v_{\rm circ}(R)$ are the particle and circular speeds at $R$, respectively,
and $\alpha \simeq 1.8$. The velocity-dependent factor provides an appropriate scaling of 
$\tau_{\rm coll}$ for eccentric orbits.

Averaging Eq. (\ref{col3}) over a 
Keplerian orbit with semimajor axis $a$ and eccentricity $e$, we obtain
\begin{equation}
 \Psi\left(a,e\right) = \left(\frac{a}{R_0}\right)^\alpha
   {\cal J}\left(e\right) \; , \label{col4}
\end{equation}
where 
\begin{equation}
 {\cal J}\left(e\right) = \frac{1}{\pi}\int_0^\pi du
  \frac{\left(1-e\cos u\right)^{1+\alpha}}{\sqrt{1+e\cos u}}\; 
 \label{col5}
\end{equation}
can be written as a series in $e^2$ with good convergence. Note that ${\cal J}\simeq 1$ for 
$e \simeq  0$, as required in Eq. (\ref{col4}), but can become $\gg$1 for very eccentric 
orbits. 

The G85's model was calibrated to match the impact fluxes of particles as measured prior
to 1985. The more recent measurements indicate lower fluxes (e.g., Dikarev 
et~al. 2005, Drolshagen et~al. 2008). Also, to estimate $\tau_{\rm coll}$, assumptions 
needed to be made in G85 about the strength of particles, and their impact speeds. As a result, 
$\tau_{\rm coll}$ proposed by G85 may have a significant uncertainty. To test 
different possibilities, we introduced two free parameters in our model, $S_1$ and $S_2$, 
that were used to shift the $\Phi(m)$ function in $\log m$ and $\log \Phi$, respectively 
(as indicated by arrows in Fig.~\ref{gf}). For example, the positive $S_2$ values 
increase $\tau_{\rm coll}$ relative to the standard G85's model, as expected, for example,
if particles were stronger than assumed in G85, or if the fluxes were lower.  

Collisional disruption of particles was taken into account during processing the output 
from the numerical integration described in Section 2.2. To account for the stochastic nature 
of breakups, we determined the break-up probability $p_{\rm coll} = 1 - \exp(-h/\tau_{\rm coll})$, 
where $h=1000\, {\rm yr}$ is the output interval, and $\tau_{\rm coll}$ was computed 
individually for each particle's orbit. The code then generated a random number $0\leq x\leq1$, 
and eliminated the particle if $x<p_{\rm coll}$. 
 
We caution that our procedure does not take into account the small debris fragments that are 
generated by disruptions of larger particles. Instead, all fragments are removed from the system.
This is an important approximation, whose validity needs to be tested in the future. 

\subsection{Model for Meteor Radar Observations}
We used the \"Opik theory (\"Opik 1951) to estimate the expected terrestrial accretion 
rate of OCC particles in our model. Wetherill (1967), and later Greenberg (1982), improved 
the theory by extending it more rigorously to the case of two eccentric orbits. Here we used 
the Fortran program written by W. F. Bottke (see, e.g., Bottke et al. 1994), which employs
the Greenberg's formalism. 

We modified the code to compute the radiants of the impacting bodies. The radiants were 
expressed in the coordinate system, where longitude $l$ was measured from the Earth's apex in 
counter-clockwise direction along the Earth's orbit, and latitude $b$ was 
measured relative to the Earth's orbital plane. Note that our definition of longitude is different 
from the one more commonly used for radar meteors, where the longitude is measured from the 
helion direction. The radiants were calculated before the effects of gravitational focusing 
were applied. 

The longitude and latitude values of radiants were binned into 1 deg$^2$ area 
segments. For each radiant bin, the code gives information about the distribution of geocentric impact 
speeds, $v_g$, and heliocentric orbits prior to the impact, as defined by $a$, $e$ and $i$. 
Here, $v_g$ is defined as $v_g = (v^2_\infty+v^2_{\rm esc})^{1/2}$, where $v_\infty$ is the relative 
velocity `in infinity' and $v_{\rm esc}$ is the escape speed from the Earth's surface. The 
orbital elements, on the other hand, are the orbital elements that the particle would have
in absence of the gravitational focusing by the Earth. The radiant distribution, $v_g$, $a$, 
$e$ and $i$ will be compared to meteor radar observations in Section 3.3. 
   
To compare our model with observations, we need to include the meteor radar detection efficiency.
This is a difficult problem because the meteor phenomenon itself and radar detection of it involve 
complex physics. For example, the specular meteor radars (SMRs), such as the Canadian Meteor Orbit Radar 
(CMOR; Campbell-Brown 2008) and Advanced Meteor Orbit Radar (AMOR; Galligan \& Baggaley 2004, 2005), 
detect the specular reflection of the meteor trail (the plasma formed by the meteoroid's passage).
The meteoroid velocities are then derived from the detection of the Fresnel diffraction patterns 
of the developing trail, or are determined by measuring the time of flight between stations. 

The detection efficiency of a meteor should mainly be a function of the particle size and 
speed, but it also depends on a number of other parameters discussed, for example, in Janches et al. 
(2008). Following W09, we opt for a simple parametrization of radar sensitivity function, where 
the detection is represented by an ionization function
\begin{equation}
I(m,v_g) = {m \over 10^{-4}\, {\rm g}} \left( {v_g \over 30\,{\rm km/s}} \right)^{3.5} \ .  
\label{cut}
\end{equation}
All meteors with $I(m,v_g)\geq I^*$ are assumed to be detected in our model, while all meteors 
with $I(m,v_g) < I^*$ are not detected (see Fentzke et al. (2009) for a similar method applied
to head echo radars). The ionization cutoff $I^*$ is different for different SMRs.
For example, $I^*\sim 1$ for CMOR (Campbell-Brown 2008) and $I^*\sim 0.001$-0.01 for AMOR (Galligan \& 
Baggaley 2004, 2005). For reference, an OCC particle with $v_g = 60$~km s$^{-1}$ and $m = 10^{-5}$~g, 
corresponding to $s\sim100$ $\mu$m, will have $I(m,v_g) \simeq 0.1$, i.e., a 
value intermediate between the two thresholds. These meteoroids would thus be detected by AMOR,
according to our definition, but not by CMOR. We will discuss these issues in more detail in 
Section~3.3.    
\section{Results}
\subsection{Orbital Evolution of OCC Particles}
\begin{figure}
\epsscale{0.48}
\plotone{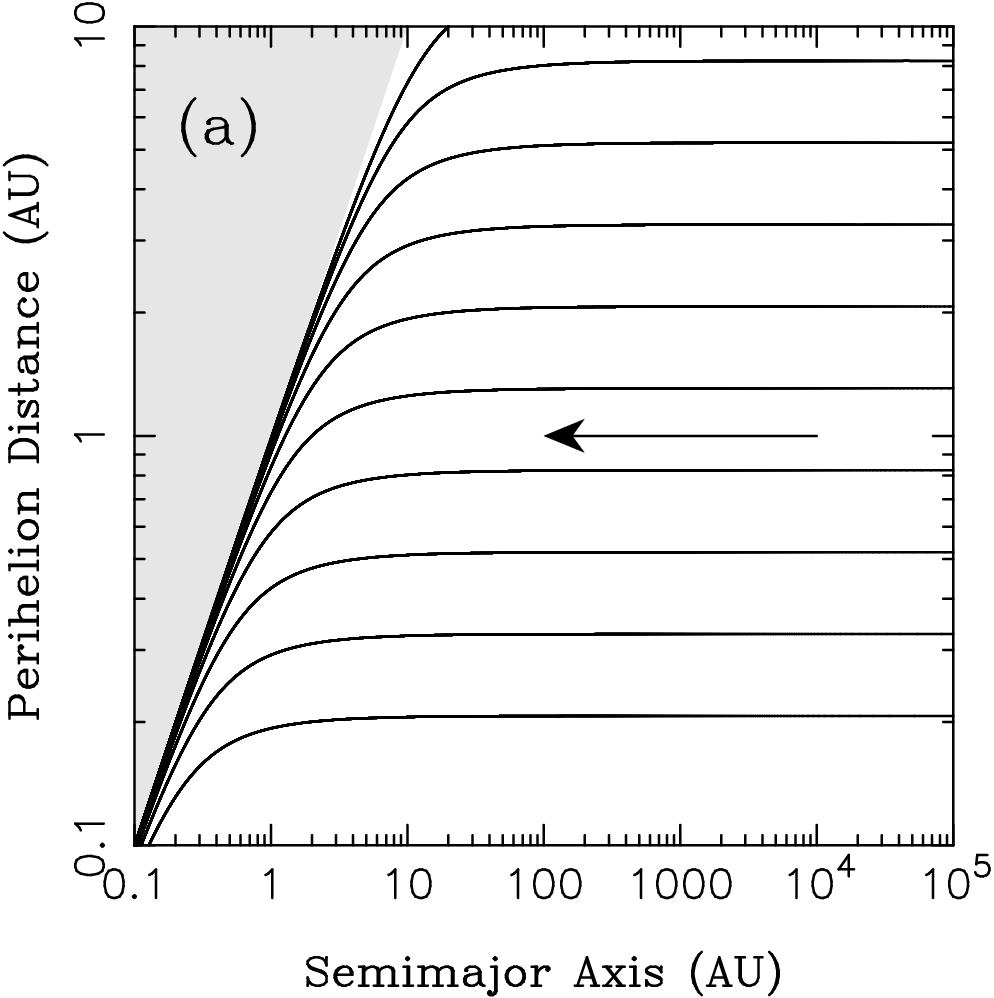}\\[5.mm]
\plotone{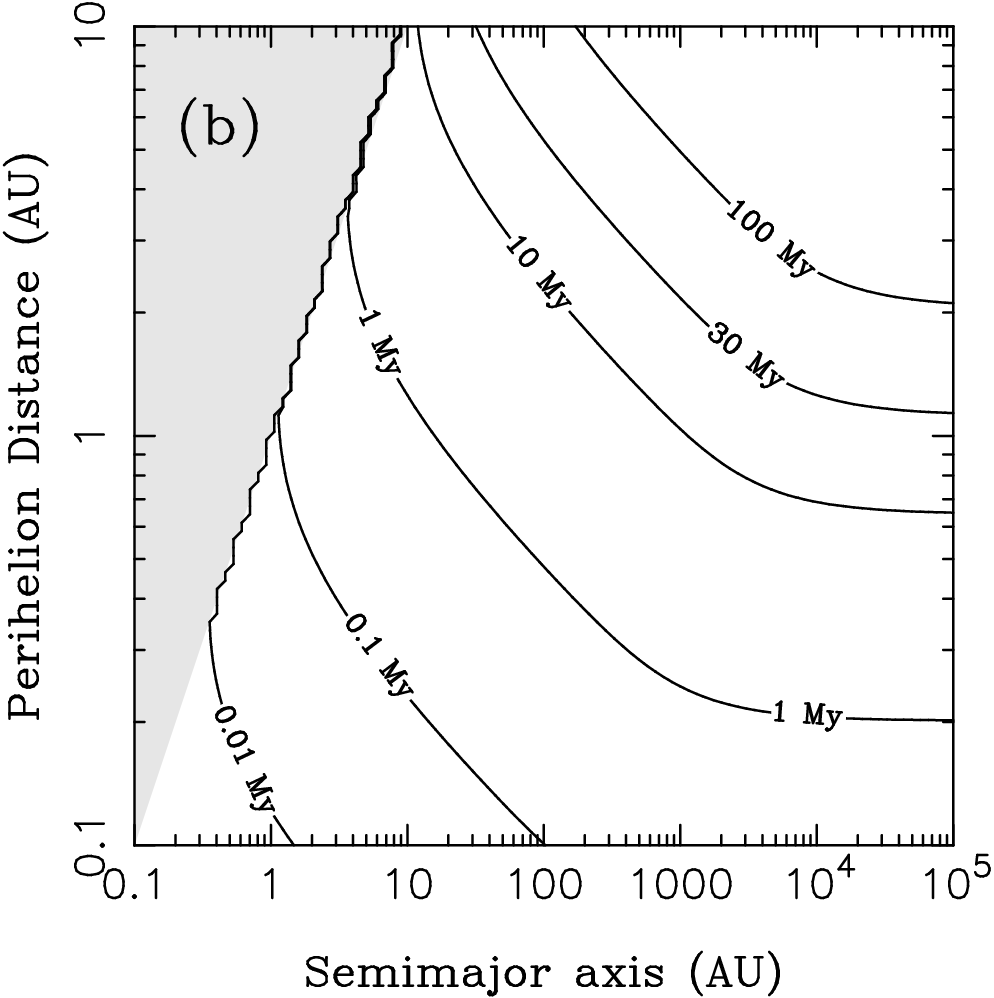}
\caption{Effects of P-R drag on the highly-eccentric orbits of OCC particles. (a) Evolution tracks of 
particles evolving by P-R drag in semimajor axis and perihelion distance. Particles with $e \simeq 1$ 
evolve from right to left along the lines of constant perihelion distance. The shaded area is inaccessible 
to orbits. (b) Evolution timescale for particles with $D=100$ $\mu$m and $\rho=2$ g cm$^{-3}$.  
Contours show the time of fall, $\tau_{\rm fall}$, from the initial orbit with $a$ and $q$ to the Sun. 
According to Eq. (\ref{tfall}), $\tau_{\rm fall}$ scales linearly with $D$ (and $\rho$), so that, 
for example, the 1 My contour for $D=100$ $\mu$m is the 10 My contour for $D=1$ mm.} 
\label{wy}
\end{figure}

Wyatt \& Whipple (1950) identified the following constant of motion of P-R drag
\begin{equation} 
C = a { (1-e^2) \over e^{4/5} }\ ,
\label{const}
\end{equation}
where $a$ and $e$ are the particle's semimajor axis and eccentricity (see also Breiter \& Jackson
1998). This constant is 
independent of the particle properties such as its size. The orbit path of any particle in $a$ and $e$ 
can thus be obtained by calculating $C$ for the initial orbit, and requiring that Eq. (\ref{const}) holds 
at all times (Fig. \ref{wy}a). For $e \sim 1$, the orbit trajectories follow the lines of 
constant $q$, because $C \simeq 2q$ for $e \simeq 1$. The semimajor axis of orbits shrinks till reaching 
a value only several times larger than $q$. At that point, a more familiar form of P-R drag takes 
place with both $a$ and $q$ converging to zero.

The timescale of orbital evolution is as follows. For an initial orbit with $a$, $e$ and 
$q=a(1-e)$, the total time of fall to the Sun is (Wyatt \& Whipple 1950)
\begin{equation}
\tau_{\rm fall} = 20\,{\rm Myr}\, \left( {s\over1\,{\rm cm}} \right)  
\left( {\rho \over 2\,{\rm g/cm^3}} \right)\left( {q \over 1\, {\rm AU}} \right)^2 f(e)\ ,
\label{tfall}
\end{equation}
where
\begin{equation}
f(e) = 1.13 {(1+e)^2 \over e^{8/5} } \int_0^e {\rm d}\eta {\eta^{3/5} \over (1-\eta^2)^{3/2}} \ .
\label{fe}
\end{equation}
Figure \ref{wy}b shows $\tau_{\rm fall}$ for orbits relevant to OCC particles.

Interestingly, the timescale for large $a$ can be relatively short if $q$ is small. For example, an 
OCC particle with $D=100$ $\mu$m, $a=10^4$ AU and $q=1$ AU takes about 20 My to fall to the Sun. 
This is only about 20 orbital periods for $a=10^4$ AU. According to Eq. (\ref{tfall}), $\tau_{\rm fall}$ 
scales linearly with particle size. Thus, a $D=10$-$\mu$m particle with the same initial orbit has 
$\tau_{\rm fall}=2$ My. 

One important aspect that cannot be captured by the analytical results discussed above is the 
effect of planetary perturbations on drifting orbits of OCC particles. To evaluate this effect, 
as described in Section 2.2, we numerically integrated the orbits of OCC particles as they evolve 
from large $a$ and interact with the planets. 

\begin{figure}
\epsscale{0.9}
\plotone{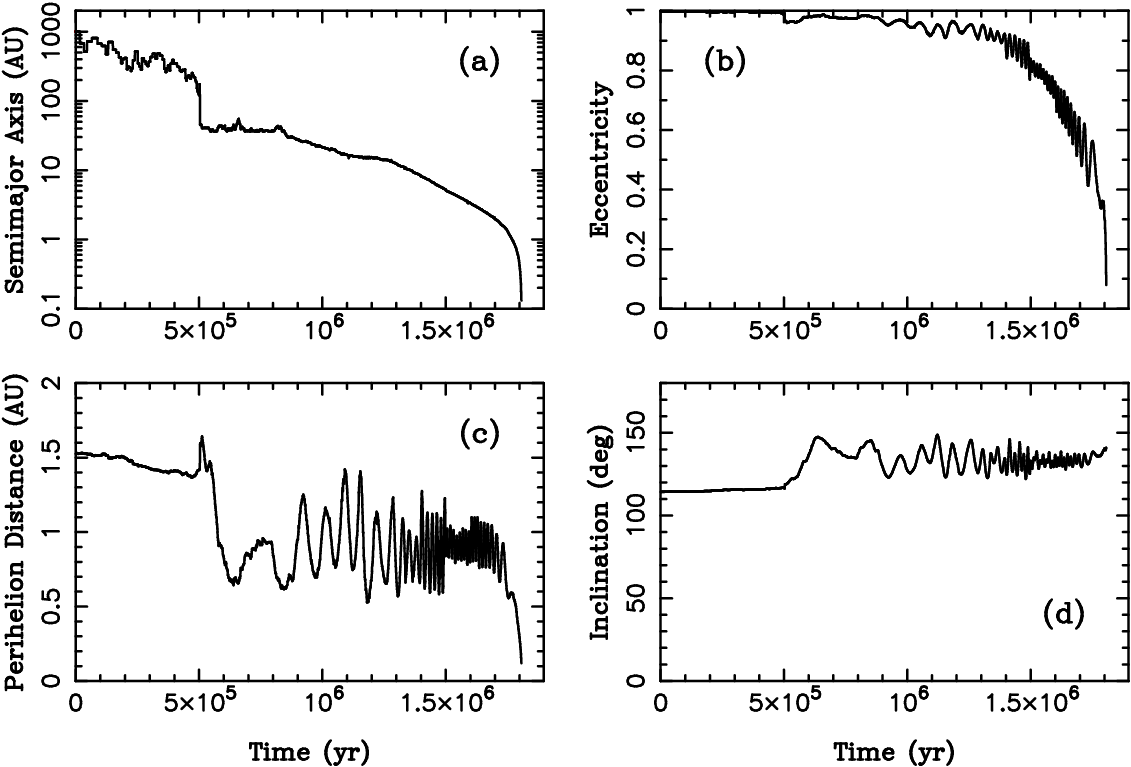}
\caption{The orbit history of a particle with $D=100$ $\mu$m, initial $a=10^3$ AU and $q=1.53$~AU. 
After 1.7 Myr, the orbit decouples from Jupiter and moves to $a\sim1$ AU. The particle ends up 
having $a<0.1$ AU, and sublimates upon reaching $R<0.05$ AU.} 
\label{orb1}
\end{figure}

Figure \ref{orb1} shows the orbital history of a particle, whose orbit evolved 
all the way down into the inner solar system. Initially, the particle's semimajor axis underwent 
a random walk caused by indirect planetary perturbations, mainly from Jupiter, during each perihelion 
passage (Wiegert \& Tremaine 1999). Then, at time $t\simeq 5\times10^5$ yr, a single perihelion passage 
produced a significant drop of $a$ from 150 to 40 AU, where the particle started to interact with 
the exterior mean motion resonances with Neptune. The following evolution was mainly controlled 
by P-R drag. Eventually, the orbit decoupled from Jupiter, reached $a\sim1$ AU, and kept shrinking 
further toward the Sun, where the particle was thermally destroyed. The large oscillations of $q$ 
for $0.9<t<1.7$ Myr, correlated with those in $i$, were produced by Kozai dynamics (e.g., Kozai 1962). 

The orbit history shown in Fig. \ref{orb1} is typical for an OCC particle that is able to make it 
into the inner solar system. These particles, however, represent a relatively small fraction of the 
initial population, with most particles being ejected from the solar system by planetary 
perturbations. Using ${\rm d}N(q)$ with $0\leq\gamma\leq1$ (Section 2.1), $a \sim 10^3$ AU and 
isotropic distribution of inclination vectors, roughly 0.8-1.5 \% of particles with $D=100$ $\mu$m 
evolve down into the inner solar system and decouple from Jupiter (as defined by the aphelion distance 
$Q=a(1+e)<4$ AU), without being disrupted by an impact (with standard $S_1=S_2=0$; Section 2.3) or 
ever having $q<0.05$ AU (to avoid sublimation). These particles can potentially be important for 
the terrestrial impact record and interpretation of the meteor radar data. The bulk of OCC particles 
that do not reach $a\sim1$ AU, do not significantly contribute to the Earth's accretion, because 
these particles spend most of their lifetimes at $R\gg1$ AU.  

The fraction of OCC particles reaching $a\sim1$ AU, $f_1$, normalized to the number of 
particles whose orbits were initially bound (see Section 2.1), is sensitive to particle size.
For $D=10$ $\mu$m, ${\rm d}N(q)$ described in Section 2.1, and initial $a \sim 10^3$ AU, 
$f_1\simeq 0.15-0.2$. For $D=300$ $\mu$m, on the other hand, $f_1 \simeq 2\times10^{-3}$. 
Moreover, for $D=1$ mm, only one particle out of the total of 5000 reached $a\sim1$ AU. 
This trend, with the larger particles having progressively smaller $f_1$ values, has 
interesting implications for the observations of sporadic meteors (Section 4).

The above estimates used $\tau_{\rm coll}$ as described in Section 2.3.2. Collisional disruption, 
however, turned out to have only a modest effect for the standard G85's $\tau_{\rm coll}$ ($S_1=S_2=0$) 
and the particle sizes considered here. For example, only $\sim$2\% of particles with 
$D=100$~$\mu$m that reached $a\sim1$ AU in our numerical integration have disrupted prior to decoupling 
from Jupiter, as detected in post-processing of the integration output, with the standard G85's 
$\tau_{\rm coll}$. A great majority of particles with $D=300$ $\mu$m also survived. The effect 
of disruptive collisions becomes more significant for $D\sim1$ mm, for which $\tau_{\rm coll}$ is 
significantly shorter (Section 2.3.2) and P-R drag is weaker. 

The thermal effects discussed in Section 2.3.1 turned out to be very important for all particle 
sizes considered here. For example, 75 out of 122 particles (i.e., over 60 \%) with $D=100$ $\mu$m 
that ever reached $Q<4$~AU, previously had $q<0.05$ AU, which is our crude threshold for the thermal 
destruction of particles. Also, 68 out of 74 $D=300$ $\mu$m particles (over 90 \%) reaching $Q<4$ 
AU previously had $q<0.05$ AU. The thermally destroyed particles are removed and the orbital 
histories after their disruption are not used for our analysis. 

These fractions are a direct consequence of the relative importance of planetary perturbations and 
P-R drag on particles with different $D$ and $q$. The orbital evolution of a large particle on the 
OCC-like orbit is primarily controlled by planetary perturbations. Sooner or later, the planets will 
eject the particle from the solar system, unless the orbit shrinks and decouples from Jupiter. 
To achieve this, $q$ of the particle's orbit must be very low, so that the P-R drag timescale 
is short (see Fig. \ref{wy}b). But if $q$ is low, it may easily drop below $q<0.05$~AU, where the 
particle is removed, thus explaining why most {\it large} particles reaching $Q<4$~AU previously have 
$q<0.05$ AU. 

Interestingly, the fraction of particles with $D\simeq 100$-300 $\mu$m reaching $a\sim1$ AU is 
not overly sensitive to the initial perihelion distance, as far as $q\lesssim5$ AU. This is because it 
is more likely to decouple if $q$ is low, because P-R drag is stronger, but this trend is nearly canceled, 
because particles with very low $q$ tend to drop below $q=0.05$ AU and sublimate before they can 
decouple. Particles with $D\simeq 100$-300 $\mu$m and $q>5$ AU, on the other hand, tend to have very 
long P-R drag timescales (e.g., $\tau_{\rm fall}>50$ Myr for initial $a=10^3$ AU; Fig. \ref{wy}b), and 
are scattered by planets from the solar system.

\subsection{Orbital and Spatial Distributions}
\begin{figure}
\epsscale{0.8}
\plotone{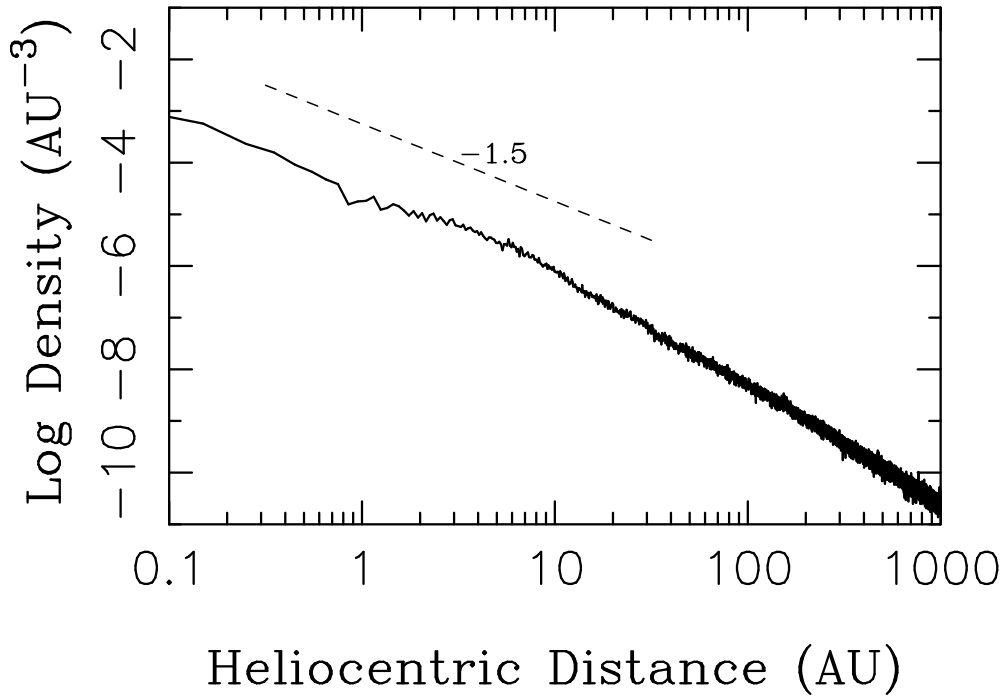}
\caption{The radial distribution of OCC particles that we obtained in our model. The number density
was normalized to the input flux of one particle released on bound orbit per year. We used 
$D=100$ $\mu$m here. The distribution with $D=300$ $\mu$m looks similar for $R>5$~AU,
but is depleted below 5 AU, relative to the one shown here, because fewer particles with $D=300$ 
$\mu$m are able to decouple from Jupiter. Particles were released on orbits with $a\sim10^3$ AU 
and uniformly random $\sin i$. The initial distribution ${\rm d}N(q)$ was set as described in Section 
2.1 with $\gamma=0.5$ (the results are not sensitive to $\gamma$).} 
\label{hist}
\end{figure}

Here we discuss the expected distribution of OCC particles in the inner solar system. Figure~\ref{hist} 
shows the number density of OCC particles as a function of $R$. As expected, the particle 
density increases toward the Sun. The radial distribution of $D=100$ $\mu$m particles can be 
approximated by a power law, ${\rm d}N(R)\propto R^{-\alpha}{\rm d}R$, where $\alpha \simeq 1.5$ for 
$R<5$ AU, and $\alpha \simeq 2.0$ for $R>10$ AU. Both these radial dependencies are significantly 
steeper than ${\rm d}N(R)\propto R^{-1}{\rm d}R$, expected for distribution of particles on 
nearly-circular orbits (see, e.g., Dermott et al. 2001). 

The relatively steep radial distribution is a consequence of Keplerian motion of particles with 
$e \sim 1$. For $R<a$, the time spent by a particle on the Keplerian orbit 
between $R$ and $R+{\rm d}R$ is ${\rm d}t \propto R^{0.5}{\rm d}R$. This leads to 
${\rm d}N(R)\propto R^{-1.5}{\rm d}R$, if the expected number of particles, which is proportional 
to ${\rm d}t$, is divided by volume $4\pi R^2 {\rm d}R$. For $R>a$, on the other hand, appropriate
for large radial distances, ${\rm d}t \sim$ const. and ${\rm d}N(R)\propto R^{-2}{\rm d}R$. 
Thus, $\alpha \simeq 1.5$-2 is expected for $e \sim 1$ (see also Liou et al. 1999).
 
We find that the number density of OCC particles at $R \sim 1$ AU is mainly contributed by the particles 
that orbitally decoupled from Jupiter. This shows the importance of orbital decoupling for the 
distribution of OCC particles in the inner solar system, and their accretion by the Earth. Specifically, 
most OCC particles accreted by the Earth are expected to have $a(1+e)<4$ AU despite the fact that 
their orbits started with $a\gtrsim10^3$ AU.    

Figure \ref{odist} shows the distributions of orbital elements for OCC particles with $D=100$~$\mu$m and 
$R<5$ AU. The semimajor axis distribution, ${\rm d}N(a)$, has a broad maximum centered at 0.5-2 AU
with $a\sim2$ AU being the most common. ${\rm d}N(a)$ decreases toward larger $a$, because particles 
with $a>3$ AU are coupled to Jupiter, have short dynamical lifetimes, and do not spend much time
at $R<5$ AU. The peak of ${\rm d}N(a)$ at $a=6$-7 AU is produced by orbits in the exterior mean motion 
resonances with Jupiter, which prolong the dynamical lifetime of particles by phase protecting 
them against encounters with Jupiter (Liou et al. 1999).

The eccentricity distribution ${\rm d}N(e)$ increases toward $e \sim 1$, which is expected because
all orbits started with $e>0.995$. The tail extending to the moderate and low eccentricity values 
is due to the dynamically long-lived particles, whose orbits decouple from Jupiter and become 
circularized by P-R drag. 

The inclination distribution is also interesting as it significantly deviates from
the initial distribution with ${\rm d}N(i) \propto \sin(i){\rm d}i$ (Fig. \ref{odist}c). 
The retrograde orbits are more common than the prograde ones. The preference for retrograde orbits  
is probably caused by gravitational perturbations from Jupiter that are more effective 
on prograde orbits, because the encounter speeds are lower, and thus $\Delta V$'s are larger. 
The prograde particles should therefore have shorter dynamical lifetimes than the retrograde 
particles, which would explain their relative paucity in a steady state ${\rm d}N(i)$ for $R<5$ AU. 
Interestingly, however, the particles that decouple from Jupiter and reach $a\sim1$ AU by P-R drag 
do not show a strong preference for retrograde orbits. 

In addition, the steady state ${\rm d}N(i)$ lacks orbits with $i\sim90^\circ$. We believe that this 
is a consequence of Kozai dynamics (e.g., Kozai 1962). It is well known that the initially near-polar 
orbits will suffer large oscillations of $q$ driven by variations of the orbital angular momentum 
vector. Most of these orbits can therefore reach very low $q$ values, where the particles will be 
destroyed by thermal effects. Indeed, by studying the orbital histories of particles that started 
with $i\sim90^\circ$, we found that most of these orbits had $q<0.05$ AU prior to reaching $Q<4$~AU. 

The steady state distribution of orbital elements of $D=300$-$\mu$m particles is similar to the one
discussed above. Instead of having a single peak at $a=6$-7 AU, however, ${\rm d}N(a)$ for $a>5$ AU is
more irregular showing many peaks and dips. Apparently, since the larger $D=300$-$\mu$m particles
drift more slowly by P-R drag, they are more susceptible to capture in a large number of resonances
(Liou et al. 1999). The second difference concerns ${\rm d}N(e)$, which is slightly more clumped toward 
$e \sim 1$ for $D=300$ $\mu$m than for $D=100$ $\mu$m. 
\subsection{Radiants and Orbits of Particles Accreted by the Earth}
The radiants of OCC particles are located near the Earth's apex (Fig. \ref{rad1}). This is logical
because the retrograde OCC particles, which come from the apex direction, have much higher 
velocity relative to the Earth ($\sim$60 km s$^{-1}$) than particles on prograde orbits. The 
retrograde particles therefore also have, according to the usual $n \sigma v$ rule, rather large 
impact probabilities with the Earth. In addition, the ionization cutoff used here (Eq. \ref{cut})
poses a rather strict limit on the mass of meteoroids that can be detected by SMRs at low speeds.
For example, a $D=100$-$\mu$m particle with $m=10^{-6}$ g and $v_g = 30$~km s$^{-1}$ has $I=0.01$,
which is near or slightly above the detection limit of AMOR, and way below the detection limit of 
CMOR. 

\begin{figure}
\epsscale{0.45}
\plotone{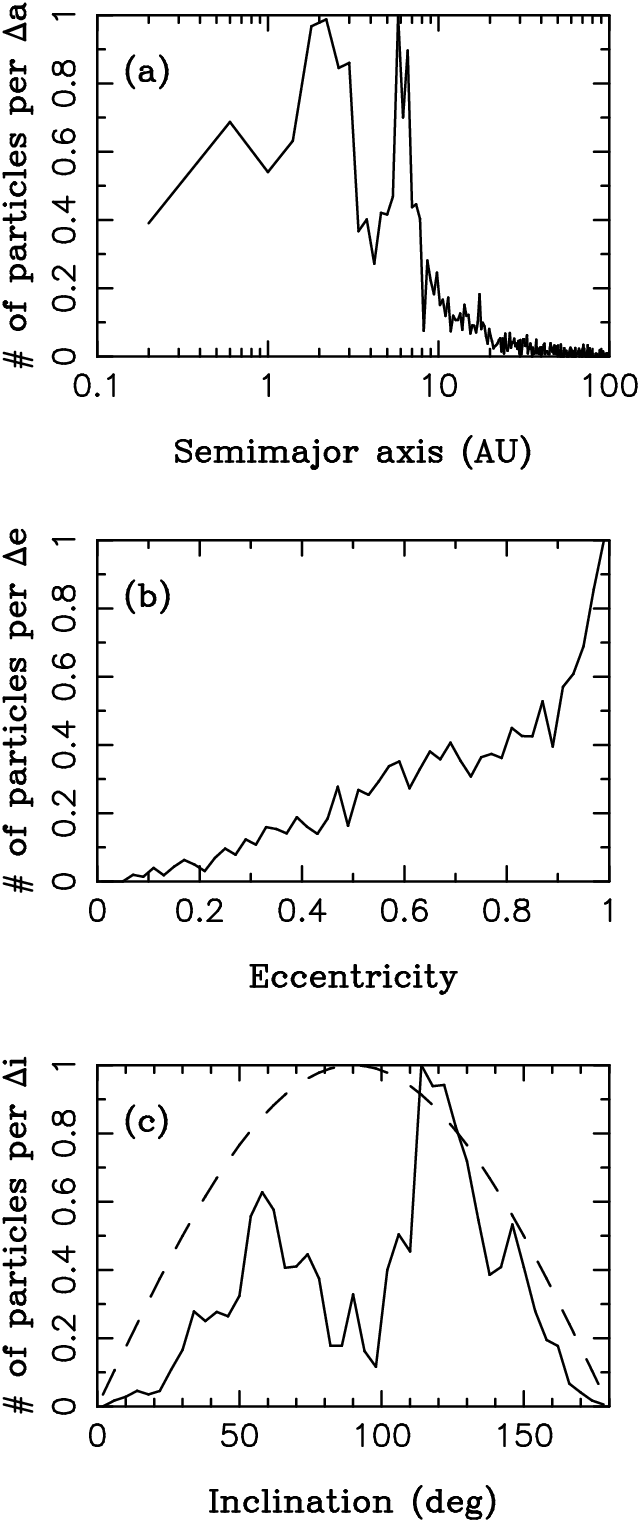}
\caption{The steady state distribution of orbital elements of particles with $D=100$~$\mu$m:
(a) ${\rm d}N(a)$, (b) ${\rm d}N(e)$, and (c) ${\rm d}N(i)$. The distributions shown here ignore 
particles with $R>5$~AU. They therefore represent the steady state distribution of orbits in the inner
solar system, which is relevant for observations of the inner zodiacal cloud and sporadic 
meteors.}  
\label{odist}
\end{figure}

\begin{figure}
\epsscale{0.7}
\plotone{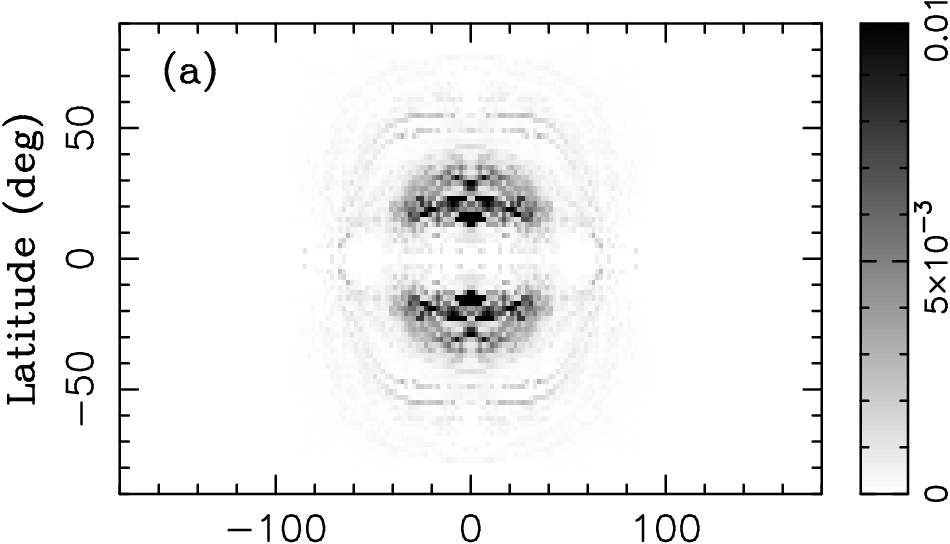}\\[5.mm]
\plotone{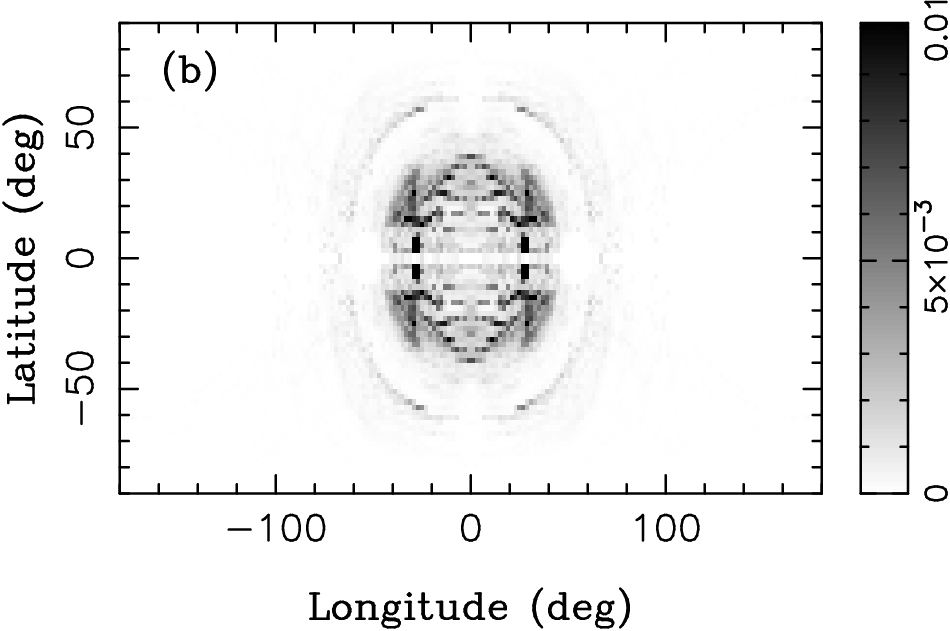}
\caption{The radiants of OCC meteoroids determined in our model with $I^*=0.01$: (a) $D=100$ $\mu$m, and
(b) $D=300$ $\mu$m. The north and south apex sources are clearly visible. In addition, the radiant 
distribution in panel (b) shows a ring structure centered at $(l,b)=(0,0)$ and extending to 
$\simeq$$60^\circ$ in $l$ and $b$. The high-frequency fluctuation of radiant density between neighbor 
bins is due to insufficient statistics and should be disregarded. The units of the side bar are 
arbitrary.} 
\label{rad1}
\end{figure}

The model radiants form the south and north apex concentrations, just as observed (e.g., Jones 
\& Brown 1993, Galligan \& Baggaley 2005, Campbell-Brown 2008). The lack of radiants within $\sim$10$^\circ$ 
about the ecliptic is due to near absence of OCC meteoroids with $i\simeq 180^\circ$ (see Fig. 
\ref{odist}c). The lack of radiants with $b>50^\circ$ (or $b<-50^\circ$) is the consequence of the 
inclination distribution shown in Fig. \ref{odist}c that is depleted in orbits with $i\sim90^\circ$.

Most north (south) radiants fall into an area on the north (south) hemisphere that has the characteristic 
triangular or half-disk shape. For $D=100$ $\mu$m, the centers of radiant concentrations are at
$b\simeq \pm20^\circ$. For $D=300$ $\mu$m, the centers are at $b\simeq \pm25^\circ$. This reflects the 
differences in ${\rm d}N(i)$ between the populations of particles with $D=100$ $\mu$m and $D=300$~$\mu$m 
that we obtained in the model. For comparison, observations indicate that $b\simeq \pm15^\circ$ 
(e.g., Galligan \& Baggaley 2005, Chau et al. 2007, Campbell-Brown 2008). In addition, the apex 
sources that we obtained in our model tend to be more stretched in both $l$ and $b$ than the observed 
ones. 

\begin{figure}
\epsscale{0.7}
\plotone{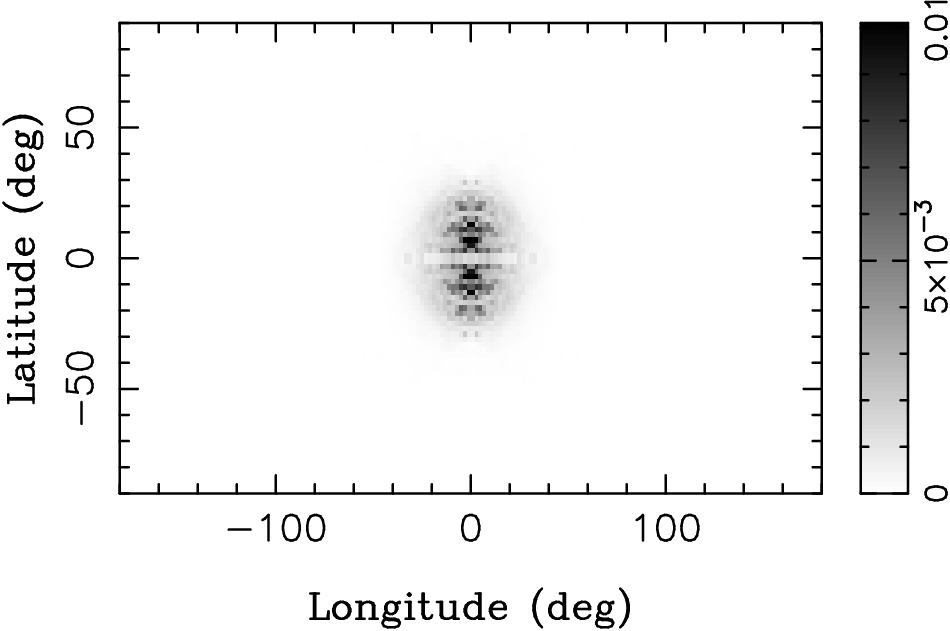}
\caption{The same as Fig. \ref{rad1} but for $D=100$-$\mu$m particles released from 1P/Halley.
The results for 55P/Tempel-Tuttle are similar.} 
\label{rad2}
\end{figure}

While more modeling work will be needed to test things with a better statistic, the issues discussed 
above may indicate that a better fit to observations could be obtained if the retrograde source had inclinations 
closer to 180$^\circ$ than the bulk of OCC particles with $100^\circ <i< 160^\circ$. It is not clear how 
this could be achieved by tweaking the parameters of our model. Instead, clues such as these seem to 
highlight the importance of known HTCs. Indeed, the two prominent active HTCs, 1P/Halley and 55P/Tempel-Tuttle, 
both have $i\simeq 162^\circ$. They would therefore be expected to produce apex sources closer to the 
ecliptic than the bulk of retrograde OCC particles (Fig. \ref{rad2}).

An interesting feature in Fig. \ref{rad1}b is the presence of a ring that stretches to $\pm60^\circ$
in longitude and latitude. A similar ring has been noted in Campbell-Brown (2008), who suggested 
that the region inside the ring can be depleted in meteor radiants, except for apex sources, because
the retrograde meteoroids with radiants inside the ring would have shorter collisional lifetimes. 
Our simple collisional model cannot reproduce this effect, because 
$\tau_{\rm coll}$ is independent of $i$ (see Section 2.3.2). In addition, W09 suggested that the 
origin of the ring can be traced back to Kozai dynamics, which confines the allowed radiants 
of particles on high-inclination orbits. Here we confirm the W09's result by isolating particles
that contribute to the ring, and checking on their orbital behavior. 

\begin{figure}
\epsscale{0.9}
\plotone{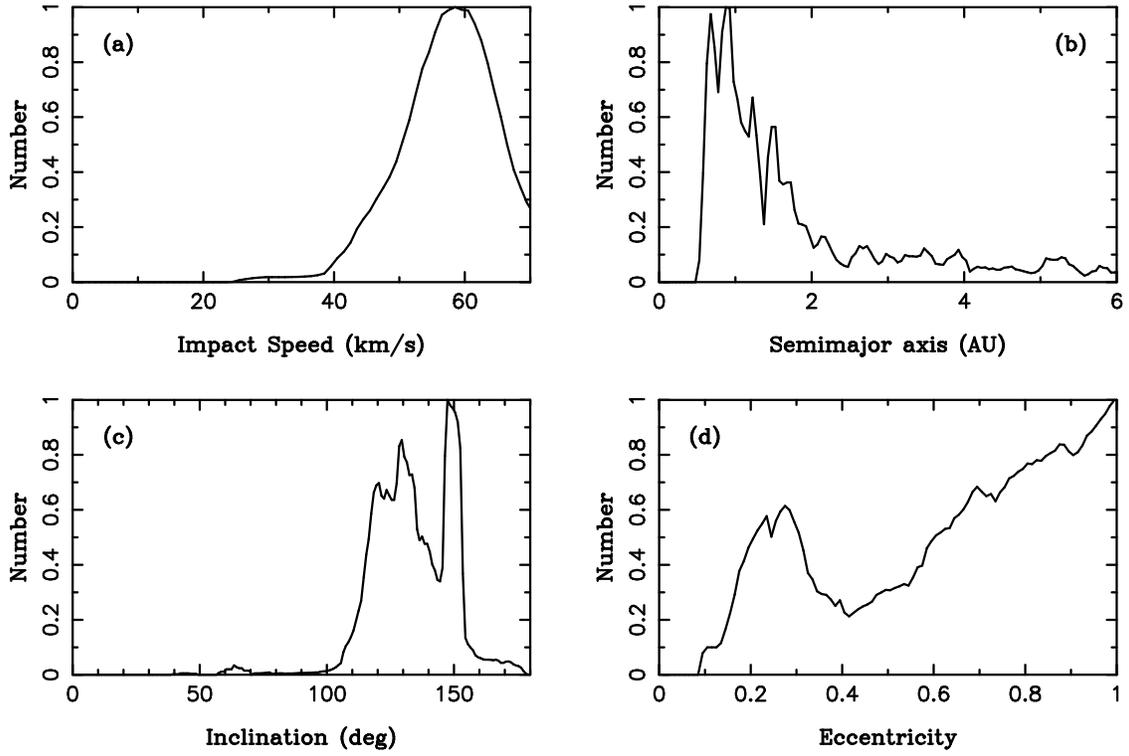}
\caption{The orbital element distribution of model apex meteoroids: (a) $v_g$, (b) $a$, (c) $i$,
and (d) $e$. Here we used OCC particles with $D=100$ $\mu$m and $I^*=0.01$. The apex meteoroids were
selected by using the following radiant cutoffs: $-40^\circ<l<40^\circ$ and $-40^\circ<b<40^\circ$.} 
\label{met11}
\end{figure}

\begin{figure}
\epsscale{0.9}
\plotone{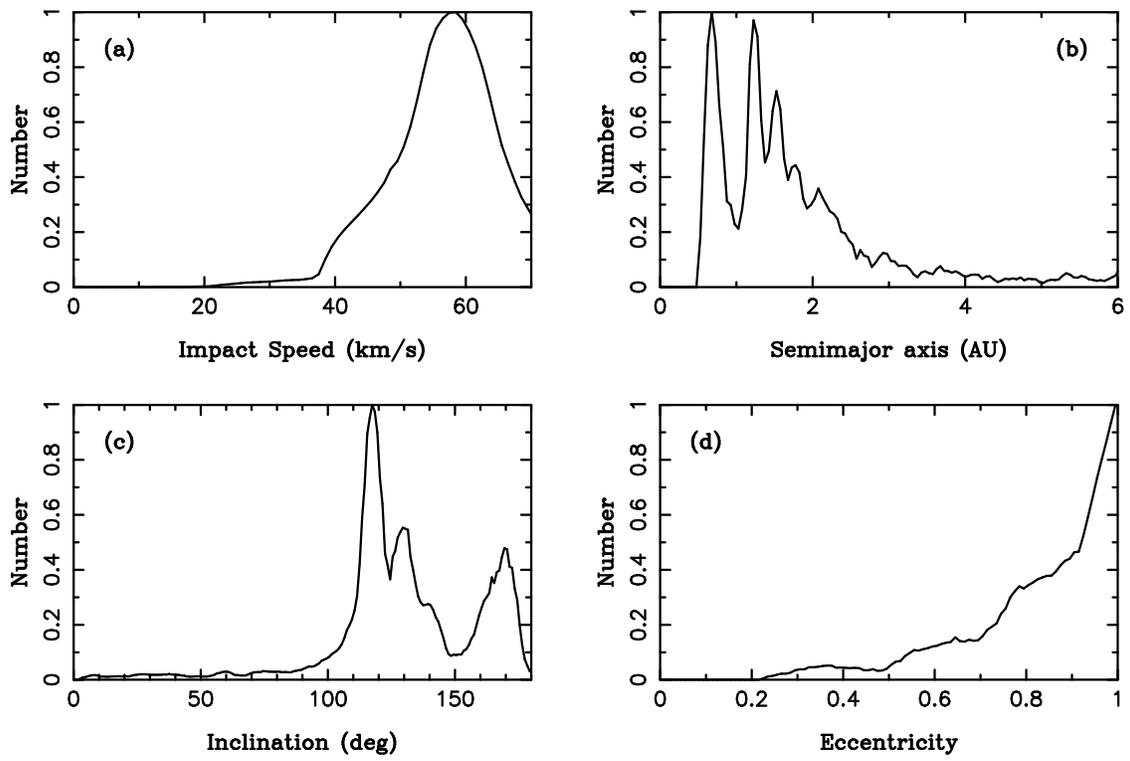}
\caption{The same as Fig. \ref{met11} but for $D=300$-$\mu$m OCC particles.} 
\label{met12}
\end{figure}

The impact speed of OCC particles peaks at $v_g \simeq  60$ km s$^{-1}$ (Fig. \ref{met11} for
$D=100$~$\mu$m and Fig. \ref{met12} for $D=300$ $\mu$m), which is a nice match to observations of apex 
meteors (Galligan \& Baggaley 2005, Campbell-Brown 2008).\footnote{Note that observations by High 
Power and Large Aperture (HPLA) radars such as AO or ALTAIR measure the apex peak speed at $\simeq55$ 
km s$^{-1}$.} The width of the peak also looks good (cf. 
Fig. 13 in Campbell-Brown 2008). In comparison, using a population of meteoroids from 1P/Halley and 
55P/Tempel-Tuttle, W09 obtained a peak at $v_g\simeq 70$ km s$^{-1}$, which is expected for large 
particles that have not evolved far from their parent comet orbit.

Figures \ref{met11} and \ref{met12} also show the model distributions of orbital elements of OCC 
meteoroids.  Distributions ${\rm d}N(a)$, now heavily weighted by the collision probability 
(cf. Fig. \ref{odist}), peak at $a\simeq 1$ AU and show a tail extending to $a>2$ AU. The 
meteoroids in the peak have orbits that have strongly evolved by P-R drag. The 
model distributions for $D=100$ $\mu$m and $D=300$ $\mu$m are similar, except for a few wiggles 
produced by statistical fluctuations. Both provide a good match to the observed semimajor axis 
distribution of apex meteors (cf. Fig. 11 in Galligan \& Baggaley 2005).

According to our model, most apex meteors should have inclinations between $i\sim100^\circ$ and 
180$^\circ$ (Figs. \ref{met11}c and \ref{met12}c), which is also the range indicated by observations. 
Unlike ${\rm d}N(i)$ measured by radars, which shows a broad peak centered at $i\sim150$-160$^\circ$, 
our model ${\rm d}N(i)$ is more spread and noisy. While part of this discrepancy could be blamed on 
insufficient statistics in our model, it may also point to a more fundamental problem. 

The eccentricity distribution is puzzling. For both $D=100$ $\mu$m and $D=300$ $\mu$m, we obtained
${\rm d}N(e)$ that raises toward $e\sim 1$. This trend is slightly more pronounced for $D=300$~$\mu$m 
(Fig. \ref{met12}d) than for $D=100$ $\mu$m (Fig. \ref{met11}d). W09, using selected HTCs for
parent bodies of apex meteors, obtained ${\rm d}N(e)$ that also peaked toward $e\sim 1$. In contrast, 
the observed apex meteors have nearly flat ${\rm d}N(e)$ at $0.2<e<1$, and show a slight depletion 
for $e<0.1$ (Fig. 13 in Campbell-Brown 2008). 

The cause of these differences is unclear. To obtain lower values of $e$ in our model, the orbits 
would need to become more circularized by P-R drag before arriving to $a\sim1$ AU. This could 
be achieved, for example, if more weight is given to particles starting with $q>1$~AU. We confirm 
this by using $\gamma>1$, but a detailed match to the observed eccentricity distribution remains 
elusive. A detailed analysis of this problem is left for future studies. 
\section{Relative Importance of Helion/Antihelion and Apex Sources}
The observations of sporadic meteors show that the relative importance of helion/anti\-helion and apex sources 
depends on the sensitivity of the radar that is used to carry out such observations. The less 
sensitive SMRs with $I^*\sim1$, such as the Harvard Radar Meteor Project (HRMP; Jones \& Brown 1993,
Taylor \& Elford 1998) or CMOR, detect $\sim$3-10 times more helion/antihelion meteors then apex meteors (see, 
e.g., Campbell-Brown (2009) for comparison of different radars). The more sensitive radars, such as 
AMOR with $I^*\sim0.001$-0.01, on the other hand, detect a relatively larger number of apex meteors. 
Finally, the apex meteors are predominant in observations by the highly-sensitive AO radar (e.g., 
Janches et al. 2003), because of their ability to detect small particles (Fentzke \& Janches 2008,
Fentzke et al. 2008).

This trend can be explained if the size frequency distribution (SFD) of apex meteoroids is steeper 
(i.e., if the number of meteoroids more sharply increases with decreasing size) than that of the 
helion/antihelion meteoroids, because radars that are capable of detecting smaller meteoroids would 
then be expected to see many more apex meteors (Fentzke \& Janches 2008).
For example, the {\it initial} SFD of particles 
produced by disrupted OCCs (or HTCs) could be steeper than the one produced by the sources 
of the helion/antihelion meteoroids (presumably active and disrupted JFCs; W09, Nesvorn\'y et 
al. 2010). While this is a possibility that cannot be ruled out by the existing data, 
there are also no indications that this might be true (McDonnell et al. 1987, Gr\"un et al. 2001,
Green et al. 2004, Reach et al. 2007).  

We propose that the predominance of apex meteors in AO observations is caused by orbital 
dynamics of particles released from OCCs (or HTCs). Let us assume that the initial SFD of 
particles released from OCCs (or HTCs) is ${\rm d} N_0(D) =N_0D^{-\zeta}{\rm d}D$. The SFD of 
meteoroids accreted by the Earth will then be
\begin{equation}
{\rm d} N(D) = f_0 P_{\rm i} N_0 D^{-\zeta} {\rm d}D  \ , 
\end{equation}
where $f_0$ is the fraction of particles that remain on bound orbits (see Section 2.1), and 
$P_{\rm i}$ is the impact probability of these particles on the Earth. Factor $f_0$ expresses 
the removal of small particles by radiation pressure (see Section 2.1). Thus, the size dependence 
of $f_0$ is such that it cannot increase the number of small particles relative to the large 
ones.\footnote{Factor $f_0(D)$ could presumably be approximated by a step function with 
$f_0(D)=0$ for $D<D^*$ and $f_0(D)=1$ for $D>D^*$, where $D^*$ is the critical diameter 
implied by Eq. (2).}

Table 1 lists $P_{\rm i}$ and $v_\infty$ for various sources. For example, the OCC particles 
with $D=30$~$\mu$m and $D=300$ $\mu$m have $P_{\rm i}=2\times10^{-6}$ and $P_{\rm i}=5\times10^{-7}$,
respectively. This indicates that ${\rm d} N(D)$ should have a {\it steeper} slope than 
${\rm d} N_0(D)$. Specifically, if ${\rm d} N_0(D)$ can be approximated by 
$D^{-\zeta}{\rm d}D$ for $D\sim30$-300 $\mu$m, where $\zeta$ is a constant, we find that ${\rm d} 
N(D) \propto D^{-(\zeta+\delta)}{\rm d}D$, where $\delta \sim 0.6$ for OCCs. This estimate was obtained 
with $0 \leq \gamma \leq 1$, $S_1\sim S_2\sim 0$, $q^*=0.05$ AU and initial $a\sim10^3$ AU.
The results for $S_2=1$, applicable if $\tau_{\rm coll}$ were $\sim$10 times longer than in G85, 
and for HTCs are similar.

For comparison, W09 suggested that comet 2P/Encke (or an orbitally similar lost comet) is the 
main source of helion/antihelion meteors. We find that particles released from comet 2P/Encke 
have $P_{\rm i}=6\times10^{-5}$ for $D=30$ $\mu$m and $P_{\rm i}=2.3\times10^{-4}$ for 
$D=300$ $\mu$m (Table~1). If these estimates are representative for the sources of the 
helion/antihelion meteors, they suggest that the slope of ${\rm d} N(D)$ should be 
{\it shallower} than that of ${\rm d} N_0(D)$ ($\delta \sim -0.6$).

We therefore find that ${\rm d} N(D)$ of apex meteoroids is expected to be steeper than that
of helion/antihelion meteoroids, even if the initial SFD of particles released from the 
respective sources --JFCs and OCCs/HTCs-- were similar. This effect is produced by orbital
dynamics of particles starting on different initial orbits. As we discussed in Section 3.1,
the Earth impact record of OCC particles is mainly contributed by those particle that decouple 
from Jupiter. Since the decoupling efficiency, described by factor $f_1$ in Section 3.1, ramps 
up toward smaller $D$, the population of small OCC (or HTC) particles is enhanced, relative to 
large ones. This effect is weaker for JFC meteoroids, for which the correlation between
$f_1$ and $P_{\rm i}$ is not as extreme. In the JFC case, the population of small particles
accreted by the Earth is suppressed by their short P-R drag timescale, and consequently,
lower $P_{\rm i}$. 

\begin{figure}
\epsscale{0.7}
\plotone{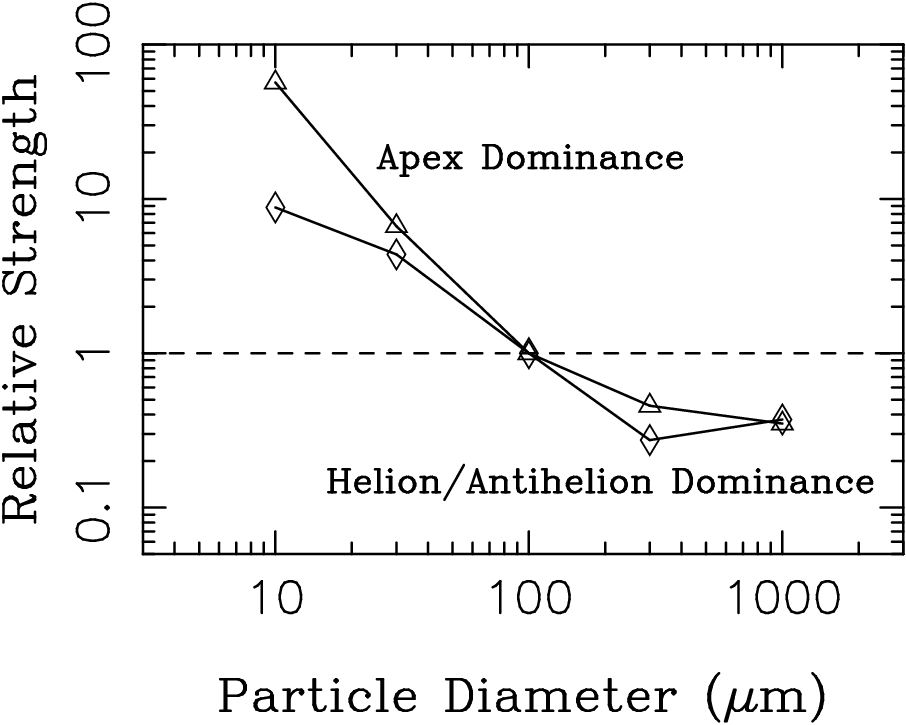}
\caption{The expected relative strength of the apex and helion/antihelion sources as a function of particle
size. The triangles (diamonds) show the result for OCC (1P/Halley) particles. We calculated the relative 
strength by dividing $P_i(D)$ listed in Table 1 for OCC (1P/Halley) particles (assumed here to represent
the apex source) by $P_i(D)$ of 2P/Encke particles (assumed to represent the helion/antihelion source; W09).
The $P_i(D)$ ratio is a proxy for the relative strength of meteor sources if ${\rm d} N_0(D)$ of different
initial populations had roughly the same shape. We arbitrarily normalized the ratio to 1 for particles
with $D=100$ $\mu$m. As more sensitive radars detect smaller particles, they are expected see more apex 
meteors, because $P_i(D)$ of OCC (1P/Halley) particles increases while that of JFC particles drops.}   
\label{ratio}
\end{figure}

The magnitude of the SFD effects discussed above is just right to explain observations. 
If the ionization threshold $I^*$ of AMOR is $\sim$100 times lower than that of CMOR/HRMP, 
these more sensitive instruments are expected to detect meteoroids that are $\sim$5 times 
smaller in size. If they detect 3-10 times more apex meteors than the helion/antihelion meteors 
(e.g., Campbell-Brown 2008), this would suggest that the SFD slope index difference between apex 
and helion/antihelion meteors is $\delta \sim 0.7$-1.4. We found $\delta \sim 1.2$ above, in a good 
agreement with observations. Figure \ref{ratio} illustrates the relative strength of sporadic 
meteor sources expected from our model. 

\section{Conclusions}
We found that only a very small fraction, $f_1\lesssim 10^{-4}$, of $D\gtrsim1$-mm OCC particles 
can ever make it into the inner solar system. The relevance of these very large OCC particles to 
observations of sporadic meteors is therefore not obvious. The situation 
looks more favorable for OCC particles with $D\simeq 100$-300 $\mu$m. These particles should 
survive the effects of radiation pressure, if released from returning OCCs at $R\gtrsim4$-12~AU. 
Moreover, about 0.2-1.5\% avoid being collisionally disrupted or thermally destroyed, decouple from 
Jupiter, and finally spiral down to $a\sim1$ AU, where their Earth-impact probability is 
increased by orders of magnitude. 

We estimated that the overall probability of Earth impact per one particle released on bound 
orbit from the returning OCC is $P_i \sim 0.5$-$1\times10^{-6}$ for $D\simeq 100$-300 $\mu$m.
This is 50-80 times lower than $P_i$ expected for particles released from HTCs such as 1P/Halley 
and 55P/Tempel-Tuttle, and 200-400 times lower than $P_i$ expected for JFCs such as 1P/Encke. 
The OCC particles will therefore significantly contribute to the sporadic meteor complex 
only if the mass of material produced by disrupting OCCs is large enough to compensate for 
these factors. From L02, we can roughly estimate that $\sim$5 returning OCCs disrupt per year 
producing the mass input of perhaps as much as $\sim10^{18}$ g yr$^{-1}$, or $3\times10^{8}$~kg~s$^{-1}$. 
Only a small fraction of this mass will end in bound particles with $D\simeq 100$-300 $\mu$m. 
For comparison, the active JFCs produce $\sim$300 kg s$^{-1}$ (Reach et al. 2007). 

We found that the SFD of apex meteoroids, presumably starting on highly-eccentric orbits, is 
expected to be steeper than those of helion/antihelion meteoroids, even if their initial SFDs
were similar. The steepening of the SFD slope of apex meteoroids results from the efficiency 
with which OCC/HTC meteoroids of different sizes decouple from Jupiter. This result has interesting 
implications for observations of sporadic meteors, because it can explain why the north/south apex 
sources are more represented in observations of highly sensitive radars that are capable of detecting 
smaller meteoroids.

\acknowledgements

This article is based on work supported by the NASA's PG\&G program. The work of DV was partially
supported by the Czech Grant Agency (grant 205/08/0064) and the Research Program MSM0021620860
of the Czech Ministry of Education. The work of DJ was partially supported by NSF Award AST 
0908118. We thank W. F. Bottke for sharing with us his \"Opik code, M.~Campbell-Brown for useful 
discussions, and Tadeusz Jopek for a very helpful review of this article.

\clearpage

\begin{table}[t]
\begin{center}
\begin{tabular}{rrr}
\hline 
$D$       & $P_{\rm i}$ & $\langle v_\infty\rangle$ \\
$\mu$m    &  $10^{-6}$ &  km s$^{-1}$ \\
\hline
\multicolumn{3}{c}{OCCs} \\
10        & 4.6 (4.6)  &  52 (52) \\ 
30        & 1.9 (1.9)  &  47 (47) \\
100       & 1.0 (1.4)  &  46 (46) \\
300       & 0.5 (0.6)  &  55 (54) \\
1000      & 0.2 (0.3)  &  58 (58) \\
\hline
\multicolumn{3}{c}{1P/Halley (HTC)} \\
10        & 57 (57)       &  60 (60)   \\ 
30        & 100 (110)     &  59 (59)  \\
100       & 80 (110)      &  59 (59)  \\
300       & 24 (46)       &  65 (61)  \\
1000      & 17 (22)       &  67 (67)  \\
\hline
\multicolumn{3}{c}{55P/Tempel-Tuttle (HTC)} \\
10        & 53 (53)   &  60 (60)  \\ 
30        & 120 (120) &  59 (59)  \\
100       & 80 (120)  &  60 (59)  \\
300       & 27 (51)   &  66 (61)  \\
1000      & 16 (23)   &  68 (66)  \\
\hline
\multicolumn{3}{c}{2P/Encke (JFC)} \\
10        & 17  (17)   &  18.0 (18.0) \\ 
30        & 60  (60)   &  18.5 (18.5) \\
100       & 210 (220)  &  18.7 (18.4) \\
300       & 230 (340)  &  23.6 (22.0) \\
1000      & 120 (300)  &  30.5 (27.0) \\
\hline
\end{tabular}
\end{center}
\caption{The Earth impact probability, $P_i$, and mean impact speed, 
$\langle v_\infty\rangle$, of particles released on different orbits. For 
each particle's diameter, $D$, we give our best estimate values for the
standard G85's $\tau_{\rm coll}$ ($S_1=S_2=0$), and for $S_1=0$ and $S_2=1$
(values in parenthesis). The longer collisional lifetime in the later case 
leads to the larger $P_i$ values. The effect of disruptive collisions is 
significant for $D\gtrsim300$ $\mu$m.}
\end{table}

\clearpage

\clearpage

\clearpage

\clearpage

\clearpage

\clearpage

\clearpage

\clearpage

\clearpage

\clearpage

\begin{thebibliography}

\bibitem[Bottke et al.(1994)]{1994Icar..107..255B} Bottke, W.~F., Nolan, 
M.~C., Greenberg, R., \& Kolvoord, R.~A.\ 1994, Icarus, 107, 255 

\bibitem[Burns et al.(1979)]{1979Icar...40....1B} Burns, J.~A., Lamy, 
P.~L., \& Soter, S.\ 1979, Icarus, 40, 1 

\bibitem[Breiter 
\& Jackson(1998)]{1998MNRAS.299..237B} Breiter, S., \& Jackson, A.~A.\ 1998, \mnras, 299, 237 

\bibitem[Campbell-Brown(2008)]{2008Icar..196..144C} Campbell-Brown, M.~D.\ 
2008, Icarus, 196, 144 

\bibitem[{\v C}apek 
\& Borovi{\v c}ka(2009)]{2009Icar..202..361C} {\v C}apek, D., \& Borovi{\v c}ka, 
J.\ 2009, Icarus, 202, 361 

\bibitem[Chau et al.(2007)]{2007Icar..188..162C} Chau, J.~L., Woodman, 
R.~F., \& Galindo, F.\ 2007, Icarus, 188, 162 

\bibitem[Choi et al.(2006)]{2006CBET..563....1C} Choi, Y.-J., Weissman, P., 
Chesley, S., Bauer, J., Stansberry, J., Tegler, S., Romanishin, W., 
\& Consolmagno, G.\ 2006, Central Bureau Electronic Telegrams, 563, 1 

\bibitem{der01} Dermott, S. F., Grogan, K., Durda, D. D., Jayaraman, S., 
 Kehoe, T. J. J., Kortenkamp, S. J., Wyatt, M. C., 2001, In Interplanetary Dust
 (Gr\"un, E., Gustafson, B.~S., Dermott, S., Fechtig, H., Eds.),
 Springer-Verlag, Berlin, 569

\bibitem[Dikarev et al.(2005)]{2005AdSpR..35.1282D} Dikarev, V., Gr{\"u}n, 
E., Baggaley, J., Galligan, D., Landgraf, M., 
\& Jehn, R.\ 2005, Advances in Space Research, 35, 1282 

\bibitem[Drolshagen et 
al.(2008)]{2008EM&P..102..191D} Drolshagen, G., Dikarev, V., Landgraf, M., 
Krag, H., \& Kuiper, W.\ 2008, Earth Moon and Planets, 102, 191 

\bibitem[Duschl et 
al.(1996)]{1996A&A...312..624D} Duschl, W.~J., Gail, H.-P., \& Tscharnuter, W.~M.\ 1996, \aap, 312, 624 

\bibitem[Dybczy{\'n}ski(2001)]{2001A&A...375..643D} Dybczy{\'n}ski, P.~A.\ 2001, \aap, 375, 643 

\bibitem[Fentzke \& Janches(2008)]{2008JGRA..11303304F} Fentzke, J.~T., \& Janches, D.\ 2008, 
Journal of Geophysical Research (Space Physics), 113, 3304 

\bibitem[Fentzke et al.(2009)]{2009JASTP..71..653F} Fentzke, J.~T., 
Janches, D., \& Sparks, J.~J.\ 2009, Journal of Atmospheric and Solar-Terrestrial Physics, 71, 653 

\bibitem[Francis(2005)]{2005ApJ...635.1348F} Francis, P.~J.\ 2005, \apj, 
635, 1348 

\bibitem[Galligan 
\& Baggaley(2004)]{2004MNRAS.353..422G} Galligan, D.~P., \& Baggaley, W.~J.\ 2004, \mnras, 353, 422 

\bibitem[Galligan 
\& Baggaley(2005)]{2005MNRAS.359..551G} Galligan, D.~P., \& Baggaley, W.~J.\ 2005, \mnras, 359, 551 

\bibitem[Green et al.(2004)]{2004JGRE..10912S04G} Green, S.~F., et al.\ 
2004, Journal of Geophysical Research (Planets), 109, 12 

\bibitem[Greenberg(1982)]{1982AJ.....87..184G} Greenberg, R.\ 1982, \aj, 
87, 184 

\bibitem[Grun et al.(1985)]{1985Icar...62..244G} Gr\"un, E., Zook, H.~A., 
Fechtig, H., \& Giese, R.~H.\ 1985, Icarus, 62, 244 

\bibitem[Gr{\"u}n et 
al.(2001)]{2001A&A...377.1098G} Gr{\"u}n, E., et al.\ 2001, \aap, 377, 1098 

\bibitem[Henning 
\& Mutschke(1997)]{1997A&A...327..743H} Henning, T., \& Mutschke, H.\ 1997, \aap, 327, 743 


\bibitem[Janches 
\& Chau(2005)]{2005JASTP..67.1196J} Janches, D., \& Chau, J.~L.\ 2005, Journal of 
Atmospheric and Solar-Terrestrial Physics, 67, 1196 

\bibitem[Janches et al.(2003)]{2003JGRA..108.1222J} Janches, D., Nolan, 
M.~C., Meisel, D.~D., Mathews, J.~D., Zhou, Q.~H., 
\& Moser, D.~E.\ 2003, Journal of Geophysical Research (Space Physics), 108, 1222 

\bibitem[Janches et al.(2006)]{2006JGRA..11107317J} Janches, D., 
Heinselman, C.~J., Chau, J.~L., Chandran, A., 
\& Woodman, R.\ 2006, Journal of Geophysical Research (Space Physics), 111, 7317 

\bibitem[Janches et al.(2008)]{2008Icar..193..105J} Janches, D., Close, S., 
\& Fentzke, J.~T.\ 2008, Icarus, 193, 105 

\bibitem[Jenniskens(2008)]{2008EM&P..102..505J} Jenniskens, P.\ 2008, Earth Moon and Planets, 102, 505 

\bibitem[Jones 
\& Brown(1993)]{1993MNRAS.265..524J} Jones, J., \& Brown, P.\ 1993, \mnras, 265, 524 

\bibitem[Jones et al.(2001)]{2001ESASP.495..575J} Jones, J., Campbell, M., 
\& Nikolova, S.\ 2001, Meteoroids 2001 Conference, 495, 575 

\bibitem[Kasuga et 
al.(2006)]{2006A&A...453L..17K} Kasuga, T., Yamamoto, T., Kimura, H., \& Watanabe, J.\ 2006, \aap, 453, L17 

\bibitem[Kessler-Silacci et al.(2007)]{2007ApJ...659..680K} 
Kessler-Silacci, J.~E., et al.\ 2007, \apj, 659, 680 

\bibitem[Kozai(1962)]{1962AJ.....67..591K} Kozai, Y.\ 1962, \aj, 67, 591 

\bibitem[Kresak(1976)]{1976BAICz..27...35K} Kresak, L.\ 1976, Bulletin of 
the Astronomical Institutes of Czechoslovakia, 27, 35 

\bibitem[Levison 
\& Duncan(1994)]{1994Icar..108...18L} Levison, H.~F., \& Duncan, M.~J.\ 1994, Icarus, 108, 18 

\bibitem[Levison et al.(2002)]{2002Sci...296.2212L} Levison, H.~F., 
Morbidelli, A., Dones, L., Jedicke, R., Wiegert, P.~A., 
\& Bottke, W.~F.\ 2002, Science, 296, 2212 

\bibitem[Liou et al.(1999)]{1999Icar..141...13L} Liou, J.-C., Zook, H.~A., 
\& Jackson, A.~A.\ 1999, Icarus, 141, 13 

\bibitem[McDonnell et 
al.(1987)]{1987A&A...187..719M} McDonnell, J.~A.~M., et al.\ 1987, \aap, 187, 719 

\bibitem[Moro-Mart{\'{\i}}n 
\& Malhotra(2002)]{2002AJ....124.2305M} Moro-Mart{\'{\i}}n, A., \& Malhotra, R.\ 2002, \aj, 124, 2305 

\bibitem[Nesvorn{\'y} et al.(2006)]{2006Icar..181..107N} Nesvorn{\'y}, D., 
Vokrouhlick{\'y}, D., Bottke, W.~F., \& Sykes, M.\ 2006, Icarus, 181, 107 

\bibitem[Nesvorn{\'y} et al.(2010)]{2010ApJ...713..816N} Nesvorn{\'y}, D., 
Jenniskens, P., Levison, H.~F., Bottke, W.~F., Vokrouhlick{\'y}, D., 
\& Gounelle, M.\ 2010, \apj, 713, 816 

\bibitem[Oort(1950)]{1950BAN....11...91O} Oort, J.~H.\ 1950, \bain, 11, 91 

\bibitem[Opik(1951)]{1951PRIA...54..165O} \"Opik, E.~J.\ 1951, Proc.~R.~Irish 
Acad.~Sect.~A, vol.~54, p.~165-199 (1951)., 54, 165 

\bibitem[Reach et al.(2007)]{2007Icar..191..298R} Reach, W.~T., Kelley, 
M.~S., \& Sykes, M.~V.\ 2007, Icarus, 191, 298 

\bibitem[Robertson(1937)]{1937MNRAS..97..423R} Robertson, H.~P.\ 1937, 
\mnras, 97, 423 

\bibitem[Steel(1996)]{1996SSRv...78..507S} Steel, D.\ 1996, \ssr, 78, 507 

\bibitem[Steel 
\& Elford(1986)]{1986MNRAS.218..185S} Steel, D.~I., \& Elford, W.~G.\ 1986, \mnras, 218, 185 

\bibitem[Taylor 
\& Elford(1998)]{1998EP&S...50..569T} Taylor, A.~D., \& Elford, W.~G.\ 1998, Earth, Planets, and Space, 50, 569 

\bibitem[Wetherill(1967)]{1967JGR....72.2429W} Wetherill, G.~W.\ 1967, 
\jgr, 72, 2429 

\bibitem[Whipple 
\& Gossner(1949)]{1949ApJ...109..380W} Whipple, F.~L., \& Gossner, J.~L.\ 1949, \apj, 109, 380 

\bibitem[Whipple(1951)]{1951ApJ...113..464W} Whipple, F.~L.\ 1951, \apj, 
113, 464 

\bibitem[Wiegert 
\& Tremaine(1999)]{1999Icar..137...84W} Wiegert, P., \& Tremaine, S.\ 1999, Icarus, 137, 84 

\bibitem[Wiegert et al.(2009)]{2009Icar..201..295W} Wiegert, P., 
Vaubaillon, J., \& Campbell-Brown, M.\ 2009, Icarus, 201, 295 

\bibitem[Wisdom 
\& Holman(1991)]{1991AJ....102.1528W} Wisdom, J., \& Holman, M.\ 1991, \aj, 102, 1528 

\bibitem[Wyatt 
\& Whipple(1950)]{1950ApJ...111..134W} Wyatt, S.~P., \& Whipple, F.~L.\ 1950, \apj, 111, 134 

\bibitem[Younger et al.(2009)]{2009MNRAS.398..350Y} Younger, J.~P., Reid, 
I.~M., Vincent, R.~A., Holdsworth, D.~A., 
\& Murphy, D.~J.\ 2009, \mnras, 398, 350 

\end{thebibliography}
\end{document}